\pdfoutput=1

\documentclass[preprint,3p]{elsarticle}

\usepackage{lineno,hyperref}
\usepackage{bm}
\usepackage{amsmath}
\usepackage{mathtools}
\usepackage{color}
\usepackage{algorithm}
\usepackage{algorithmic}
\usepackage{multirow}
\modulolinenumbers[5]

\usepackage{amssymb}

\usepackage{subfigure}
\usepackage{array}
\newcolumntype{C}[1]{>{\centering\arraybackslash}p{#1}}

\usepackage{bigints}

\usepackage{bbding}

\usepackage{units}

\usepackage{xcolor}
\usepackage{graphicx,import}

\newcommand{\Div}[1]{\nabla \cdot {#1}}
\newcommand{\Grad}[1]{\nabla {#1}}

\usepackage{stmaryrd} % für sprungoperator

\newcommand{\avg}[1]{\{\!\{#1\}\!\}}

\newcommand{\jump}[1]{\llbracket {#1} \rrbracket }

\newcommand{\intele}[2]{ \left( {#1},{#2} \right)_{\Omega_{e}} }
\newcommand{\inteleface}[2]{ \left( {#1},{#2} \right)_{\partial\Omega_{e}} }
\newcommand{\intelefaceInterior}[2]{ \left( {#1},{#2} \right)_{\partial\Omega_{e}\setminus\Gamma_h }}

\newenvironment{remark}[1][Remark]{\begin{trivlist}
\item[\hskip \labelsep {\bfseries #1}]}{\end{trivlist}}

\usepackage{enumitem}
\setlist[enumerate]{label*=\roman*),ref=\roman*)}

\journal{Journal}

\begin{document}

\begin{frontmatter}

\title{Modern discontinuous Galerkin methods for the simulation of\\ transitional and turbulent flows in biomedical engineering:\\ A comprehensive LES study of the FDA benchmark nozzle model}

\author{Niklas Fehn}
\ead{fehn@lnm.mw.tum.de}
\author{Wolfgang A. Wall}
\ead{wall@lnm.mw.tum.de}
\author{Martin Kronbichler\corref{correspondingauthor1}}
\cortext[correspondingauthor1]{Corresponding author at: Institute for Computational Mechanics, Technical University of Munich, Boltzmannstr. 15, 85748 Garching, Germany. Tel.: +49 89 28915300; fax: +49 89 28915301}
\ead{kronbichler@lnm.mw.tum.de}
\address{Institute for Computational Mechanics, Technical University of Munich,\\ Boltzmannstr. 15, 85748 Garching, Germany}

\begin{abstract}
This work uses high-order discontinuous Galerkin discretization techniques as a generic, parameter-free, and reliable tool to accurately predict transitional and turbulent flows through medical devices. Flows through medical devices are characterized by moderate Reynolds numbers and typically involve different flow regimes such as laminar, transitional, and turbulent flows. Previous studies for the FDA benchmark nozzle model revealed limitations of Reynolds-averaged Navier--Stokes turbulence models when applied to more complex flow scenarios. Recent works based on large-eddy simulation approaches indicate that these limitations can be overcome but also highlight potential limitations due to a high sensitivity with respect to numerical parameters. The novel methodology presented in this work is based on two key ingredients. Firstly, we use high-order discontinuous Galerkin methods for discretization in space yielding a discretization approach that is robust, accurate, and generic. The inherent dissipation properties of high-order discontinuous Galerkin discretizations render this approach well-suited for transitional and turbulent flow simulations. Secondly, a precursor simulation approach is applied in order to correctly predict the inflow boundary condition for the whole range of laminar, transitional, and turbulent flow regimes. This approach eliminates the need to fit parameters of the numerical solution approach. We investigate the whole range of Reynolds numbers as suggested by the FDA benchmark nozzle problem in order to critically assess the predictive capabilities of the solver. The results presented in this study are compared to experimental data obtained by particle image velocimetry demonstrating that the approach is capable of correctly predicting the flow for different flow regimes.
\end{abstract}


\end{frontmatter}

\section{Motivation}\label{Motivation}
Although Reynolds-averaged Navier--Stokes solvers are currently the state-of-the-art tool in industrial design, there is a strong need for more reliable and accurate flow solvers. This need has been exemplified by the FDA benchmark nozzle problem, a benchmark initiated by the U.S. Food and Drug Administration (FDA) in order to assess the state-of-the-art of CFD methods for the design of medical devices~\cite{Malinauskas2017}. This test case is particularly interesting and challenging from a numerical point of view since the Reynolds number of this benchmark problem has been chosen in a way that all the regimes of laminar, transitional, and turbulent flows are covered, which is typical of medical device flow problems. The FDA benchmark nozzle model has been investigated experimentally in~\cite{Hariharan2011,Raben2016} as well as numerically using Reynolds-averaged Navier--Stokes (RANS) simulation approaches in~\cite{Stewart2012,Stewart2013,Bhushan2013} and large-eddy simulation (LES) approaches in~\cite{Delorme2013,Passerini2013,Janiga2014,Chabannes2017,Zmijanovic2017,Nicoud2018}. A hybrid RANS/LES approach has been analyzed in~\cite{Bhushan2013}. Results based on the RANS strategy showed major discrepancies compared to experimental results revealing the limitations of this turbulence modeling approach especially if the problem under consideration involves more complex flow features as well as different flow regimes such as laminar and turbulent flows.
\begin{table}
\caption{Previous LES studies for FDA benchmark nozzle problem characterized in terms of the considered range of throat Reynolds numbers, the spatial discretization approach, the turbulence modeling strategy, and the type of inflow boundary conditions (FD: finite difference, FE: finite element, FV: finite volume, DG: discontinuous Galerkin, IBM: immersed boundary method, TI: turbulence injection). The symbol \XSolidBrush means that the respective Reynolds number has not been considered, (\Checkmark) means that results have been shown in a qualitative manner (contour plots) or in quantitative manner but without a critical assessment of the results. The symbol \Checkmark implies that a detailed numerical investigation of the respective test case has been performed and that the results have been assessed critically, e.g., in terms of a mesh refinement study or other measures demonstrating the robustness and reliability of the results.}
\label{tab:PreviousLES}
\renewcommand{\arraystretch}{1.1}
\begin{center}
\begin{scriptsize}
\begin{tabular}{lllllllll}
\hline
Study & \multicolumn{5}{l}{Throat Reynolds number~$\mathrm{Re}_{\mathrm{th}}$} & Spatial discretization & Turbulence model & inflow BC\\
\cline{2-6}
& 500 & 2000 & 3500 & 5000 & 6500 & & &\\
\hline
Delorme et al.~\cite{Delorme2013}       & (\Checkmark) & (\Checkmark) & (\Checkmark) & (\Checkmark) & \XSolidBrush & high-order FD, IBM & Vreman & parabolic\\
% Delorme: results are only shown for a single mesh, authors mention that a grid independence study has been performed for the transitional case Re=2000 ("Indeed, grid independence is achieved for the transitional case", p. 14, bottom left)
Passerini et al.~\cite{Passerini2013}   & \Checkmark & (\Checkmark) & (\Checkmark) & \XSolidBrush & \XSolidBrush & low-order FE & no model & parabolic\\
% Passerini: mesh convergence study for Re=500, no mesh convergence study for Re=2000 and 3500, Reynolds numbers larger than 3500 have not been considered due to limitations in computational costs
Janiga~\cite{Janiga2014}                & \XSolidBrush & \XSolidBrush & \XSolidBrush & \XSolidBrush & (\Checkmark) & low-order FV & Smagorinsky & parabolic\\
% Janiga: no mesh convergence study shown
Chabannes et al.~\cite{Chabannes2017}   & \Checkmark & (\Checkmark) & (\Checkmark) & \XSolidBrush & \XSolidBrush & Taylor-Hood FE & no model & parabolic\\
Zmijanovic et al.~\cite{Zmijanovic2017}& (\Checkmark) & \XSolidBrush & \Checkmark & \XSolidBrush & \XSolidBrush & fourth order FV & Sigma & parabolic, TI\\
% Zmijanovic: different mesh refinements investigated but no grid-converged results could be shown (also for the "robust" turbulence injection approach, the medium and fine mesh still show comparably large differences even though they are way smaller compared to the case without fluctuations at the inflow)
Nicoud et al.~\cite{Nicoud2018} 	   & (\Checkmark) & (\Checkmark) & \Checkmark & (\Checkmark) & \XSolidBrush & fourth order FV & Sigma & parabolic, TI\\
present study                          & \Checkmark & \Checkmark & \Checkmark & \Checkmark & \Checkmark & high-order DG & no model & precursor\\
\hline
\end{tabular}
\end{scriptsize}
\end{center}
\renewcommand{\arraystretch}{1}
\end{table}
Computationally more expensive LES studies have been performed in order to explain differences between experimental results and RANS simulation results. In contrast to the RANS approach, large-eddy simulation has the advantage that the influence of turbulence modeling tends to zero for increasing mesh resolution, finally reproducing the results of direct numerical simulation (DNS) results for sufficiently fine mesh resolutions.
In previous LES studies, different spatial discretization techniques such as finite volume, finite difference, and finite element methods have been used. In the context of LES subgrid-scale modeling, different turbulence modeling strategies have been applied in these studies. These modeling approaches include both physical subgrid scale models as well as implicit/no-model LES strategies. An overview of previous LES studies for the FDA benchmark nozzle problem is given in Table~\ref{tab:PreviousLES}, where the different studies are characterized in terms of the range of considered throat Reynolds numbers, the numerical discretization scheme, the turbulence modeling strategy, and the type of inflow boundary conditions. In these works good agreement with experimental results have been shown for specific Reynolds numbers and mesh resolutions.
However, there are many open issues and shortcomings motivating to further analyze this test case in the present work. Several aspects are worth mentioning: None of the previous LES studies considered the full range of throat Reynolds numbers~$\mathrm{Re}_t=500-6500$ as suggested by the initiators of the FDA benchmark in order to assess the predictive capabilities of a flow solver applied to different flow regimes. Only three of the six previous LES studies have shown mesh refinement studies. Mesh refinement studies with different meshes are shown in~\cite{Passerini2013,Chabannes2017} for the laminar test case at~$\mathrm{Re}_{\mathrm{th}}=500$ and in~\cite{Zmijanovic2017} for the turbulent test case at~$\mathrm{Re}_{\mathrm{th}}=3500$. All previous LES studies prescribed a parabolic velocity profile at the inlet, although the flow might be transitional or turbulent at the inlet for higher Reynolds numbers. It is well-known that inflow boundary conditions can have a significant impact on the numerical results especially for transitional and turbulent flow problems. Previous studies~\cite{Passerini2013,Zmijanovic2017} mentioned that the numerical results obtained for the jet breakdown location have been very sensitive to a change in the numerical parameters. This aspect has been addressed critically in the recent work~\cite{Zmijanovic2017} for the~$\mathrm{Re}_{\mathrm{th}}=3500$ test case, where turbulent fluctuations have been added to the laminar parabolic inflow profile in order to reduce the sensitivity of the numerical results with respect to the jet breakdown location. However, it remains unclear why this robust strategy has not been applied to the transitional test case at~$\mathrm{Re}_{\mathrm{th}}=2000$, which can be expected to be even more sensitive with respect to the jet breakdown location. In~\cite{Nicoud2018}, results for the transitional test case at~$\mathrm{Re}_{\mathrm{th}}=2000$ and the turbulent test case at~$\mathrm{Re}_{\mathrm{th}}=5000$ are shown in addition to the results first presented in~\cite{Zmijanovic2017}, but again results are only shown for one specific spatial resolution. Let us emphasize that the original FDA benchmark has been performed in a blinded way, i.e., the 28 participating CFD groups submitted numerical results without having knowledge of experimental results. More importantly, the FDA initiative raises the question regarding the predictive capabilities of state-of-the-art CFD methods for the simulation of transitional and turbulent flows. In our opinion, demonstrating reproducibility is only the first step towards answering this question. As exemplified in~\cite{Zmijanovic2017}, excellent agreement with experimental results might easily be achieved for a specific set of parameters. However, it remains unclear whether LES results are really converged and robust if results are only shown for one spatial resolution. To explicitly address this aspect and given the fact that experimental results are already published for this problem, we show LES results including mesh refinement studies for all Reynolds numbers. Moreover, it appears to be a necessity to use the same setup in terms of geometry, boundary conditions, numerical discretization parameters, etc.~for all Reynolds numbers. Compared to previous LES studies performed for this benchmark problem, the present work is characterized mainly by two distinctive features:

\begin{itemize}
\item We use high-order discontinuous Galerkin methods for discretization in space without explicit subgrid-scale model. High-order DG methods are well-suited for implicit large-eddy simulation as demonstrated in~\cite{Uranga2011,Gassner2013,Beck2014b,Wiart2014,Wiart2015,Fernandez2017} due to the inherent dissipation properties of discontinuous Galerkin schemes. This approach, sometimes referred to as under-resolved DNS, leads to a discretization scheme that is parameter-free and generic in the sense that it can be successfully applied to laminar, transitional, and turbulent flows without the need to adapt the numerical discretization scheme. For example, physical subgrid-scale models suffer from the fact that they cannot correctly identify the flow regime, and therefore also introduce a turbulent viscosity in laminar flow situations. There is an ongoing debate in the literature whether and to which extent an implicit LES approach based on high-order DG discretizations can benefit from the use of explicit subgrid-scale models in the limit of large Reynolds numbers~\cite{VanDerBos2010,Beck2016,Chapelier2016,Flad2017,Manzanero2018}, but it appears to be widely accepted that such a high-order DG discretization approach without explicit modeling is hard to beat in the transitional and low Reynolds number turbulent regime. Since this benchmark problem covers both the laminar and turbulent flow regime, choosing a generic flow solver without explicit LES modeling is an important aspect in our opinion.

\item We use a precursor simulation approach in order to prescribe the velocity field at the inflow boundary. Basically, synthesis methods as well as precursor simulation techniques can be used to generate turbulent inflow boundary conditions, see for example the comparative study~\cite{Keating2004}, the review article~\cite{Tabor2010}, and the recent works~\cite{Shur2014,Schmidt2017} on synthetic turbulence generators. The precursor simulation approach is used in this work for the following reasons: Firstly, synthesis methods require knowledge about the flow under consideration (laminar or turbulent flow regime at the inflow, turbulence intensities and length/time scales at the inflow boundary). Since this information is not available for the FDA test case analyzed here, using a synthesis approach would probably end up in a parameter fitting in our opinion, i.e., the simulations would have to be performed several times for different parameters in order to match experimental results which is against the spirit of this benchmark problem. Secondly, the precursor approach is generic, i.e., it can be expected to provide physically correct inflow data independently of the flow regime. The aim of the present study is to assess numerical methods for the simulation of turbulent flows and their ability to correctly predict the flow behavior in the laminar, transitional, and turbulent regimes without introducing knowledge about the flow (e.g., from experimental results) into the simulation setup.
\end{itemize}

Due to the high accuracy of discontinuous Galerkin spectral element methods in combination with a computationally efficient matrix-free implementation we are able to rigorously analyze this benchmark problem for all Reynolds numbers allowing significant progress compared to previous numerical studies.

\medskip
The outline of this article is as follows. In Section~\ref{NumericalDiscretization} we summarize the numerical discretization techniques used to solve the incompressible Navier--Stokes equations with a focus on the description of the discontinuous Galerkin discretization as a generic approach for the simulation of laminar and turbulent flows. The FDA benchmark nozzle test case is summarized in Section~\ref{TestCase} where we discuss aspects related to the geometry, the boundary conditions as well as the precursor simulation approach, and the mesh used for the numerical investigations. In Section~\ref{NumericalResults}, numerical results are presented for Reynolds numbers in the range~$\mathrm{Re}_{\mathrm{th}}=500-6500$ covering the laminar, transitional, and turbulent regime.
%Additionally, we suggest to extend this benchmark by another Reynolds number of~$\mathrm{Re}_{\mathrm{th}}=8000$.
A discussion of the results as well as our main conclusions close this article in Section~\ref{Conclusion}.

\section{Numerical discretization techniques}\label{NumericalDiscretization}
This work targets the numerical solution of the incompressible Navier--Stokes equations given as
\begin{align}
\frac{\partial \bm{u}}{\partial t} + \nabla \cdot \bm{F}_{\mathrm{c}}(\bm{u}) - \nabla \cdot \bm{F}_{\mathrm{v}} (\bm{u}) + \nabla p &= \bm{f} \; ,\label{MomentumEquation}\\
\nabla \cdot \bm{u} &= 0 \; ,\label{ContinuityEquation}
\end{align}
which are supplemented by initial and boundary conditions to obtain a well-defined problem. Here,~$\bm{u}$ denotes the velocity vector in~$d$ space dimensions and~$p$ the kinematic pressure. The convective term is given as~$\bm{F}_{\mathrm{c}}(\bm{u}) = \bm{u}\otimes \bm{u}$ and the viscous term as~$\bm{F}_{\mathrm{v}} (\bm{u}) = \nu \Grad{\bm{u}}$ with the kinematic viscosity~$\nu$. The body force term on the right-hand side of the momentum equation is denoted by~$\bm{f}$. To solve these equations numerically, the incompressible Navier--Stokes equations are discretized by projection methods in time and by high-order discontinuous Galerkin methods in space, as detailed in the following Sections~\ref{TemporalDiscretization} and~\ref{SpatialDiscretization}, respectively.

\subsection{Temporal discretization: Projection methods as efficient solution strategies}\label{TemporalDiscretization}
Projection methods have a long tradition in solving the incompressible Navier--Stokes equations. Instead of solving a monolithic system of equations for both velocity and pressure unknowns, a set of easier-to-solve equations is obtained by splitting the velocity and pressure unknowns. A pressure Poisson equation is derived by inserting the momentum equation into the continuity equation and a divergence-free velocity field is obtained by applying a projection onto the space of divergence-free vectors. For the velocity unknowns, a (convection--)diffusion problem has to be solved depending on the temporal treatment of the convective term. The first projection method has been proposed by Chorin~\cite{Chorin1968} in 1968. Since this splitting scheme suffers from low-order accuracy in time, projection methods achieving higher order accuracy in time have subsequently been developed, see for example~\cite{VanKan1986,Timmermans1996,Karniadakis1991} and the review article~\cite{Guermond2006}. Projection methods are widely used due to their algorithmic simplicity as well as their computational efficiency especially for high-Reynolds number flows~\cite{Karniadakis2013}.

In the present study, we apply the incremental pressure-correction scheme in rotational formulation~\cite{Guermond2006} which is formally second order accurate in the~$L^2$-norm of the velocity. The solution of each time step consists of the following substeps
\begin{align}
\frac{\frac{3}{2} \hat{\bm{u}}-2\bm{u}^{n}+\frac{1}{2}\bm{u}^{n-1}}{\Delta t}
+ \Div{\bm{F}_{\mathrm{c}}\left(\hat{\bm{u}}\right)} -\Div{\bm{F}_{\mathrm{v}}\left(\hat{\bm{u}}\right)} &=
- \Grad{p^{n}}
+ \bm{f}\left(t_{n+1}\right)\; ,\label{PressureCorrection_MomentumStep}\\
-\nabla^2 \phi^{n+1} &= -\frac{3}{2 \Delta t}\Div{\hat{\bm{u}}} \; ,\label{PressureCorrection_PressurePoissonEquation}\\
p^{n+1} &= \phi^{n+1} + p^{n} - \nu \Div{\hat{\bm{u}}}\; ,\label{PressureCorrection_PressureUpdate}\\
\bm{u}^{n+1} &= \hat{\bm{u}} - \frac{2 \Delta t}{3} \Grad{\phi^{n+1}}\; .\label{PressureCorrection_Projection}
\end{align}
An implicit treatment of the viscous term as well as the convective term is used in the momentum equation~\eqref{PressureCorrection_MomentumStep} to avoid restrictions of the time step size, e.g., according to the Courant--Friedrichs-Lewy (CFL) condition. In this first substep, an intermediate velocity~$\hat{\bm{u}}$ is computed. A Poisson equation is solved for the pressure increment~$\phi$ in equation~\eqref{PressureCorrection_PressurePoissonEquation}, and the pressure is updated in equation~\eqref{PressureCorrection_PressureUpdate}. Finally, a divergence-free velocity field is calculated by solving the projection equation~\eqref{PressureCorrection_Projection}.

\subsection{Spatial discretization: High-order discontinuous Galerkin methods}\label{SpatialDiscretization}
The development of discontinuous Galerkin discretization techniques for the numerical solution of the Navier--Stokes equations is currently the subject of intensive research. The great potential of discontinuous Galerkin methods originates from the fact that this approach allows to combine advantages of both finite volume methods and finite element methods~\cite{Hesthaven2007}. On the one hand, high-order accurate discretization approaches can be constructed by expressing the numerical solution as high-order polynomial approximations inside each element similar to continuous finite element methods. The construction of high-order methods can be seen as a main limitation of finite volume methods. On the other hand, discontinuous Galerkin methods rely on the concept of numerical fluxes having their origin and being established in the finite volume community. By the use of suitable numerical flux functions, discretization schemes with good stability properties for convection-dominated problems can be constructed. In other words, finite volume methods can be interpreted as a low-order variant of the discontinuous Galerkin method.

\begin{figure}[!ht]
\centering
  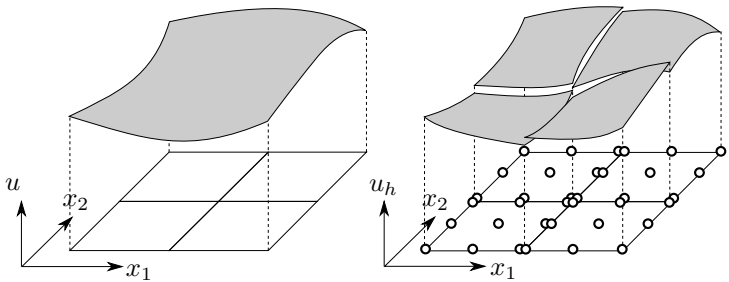
\caption{Approximation of exact solution~$u$ by a piecewise smooth numerical solution~$u_h$ with discontinuities between elements. Inside each element the solution is approximated by a polynomial basis using Lagrange polynomials, where the nodal points shown on the right-hand side would correspond to shape functions of polynomial degree~$k=2$.}
\label{fig:DiscontinuousSpaces}
\end{figure}

To approximate the exact solution~$u$ by a discrete numerical solution~$u_h$, the computational domain is subdivided into elements,~$\Omega_h = \bigcup_{e=1}^{N_{\text{el}}} \Omega_{e}$. In case of discontinuous Galerkin approaches the numerical solution exhibits discontinuities between elements while the solution is typically approximated by polynomial functions inside each element
\begin{align}
u_h^e(\bm{x}(\bm{\xi})) = \sum_{e=1}^{N_{\mathrm{DoFs},e}} N_i \left(\bm{\xi}\right) u_i^e \ ,
\end{align}
where~$u_i^e$ are the~$N_{\mathrm{DoFs},e}$ unknown solution coefficients on element~$e$. The argument~$\bm{\xi}$ expresses that the shape functions~$N_i^e (\bm{\xi})$ are defined on a reference element with coordinates~$\bm{\xi} \in \left[0,1\right]^d$, where a high-order polynomial mapping~$\bm{x}(\bm{\xi})$ is used for the transformation from reference space to physical space. This basic idea is illustrated in Figure~\ref{fig:DiscontinuousSpaces} for the two-dimensional case where the mapping~$\bm{x}(\bm{\xi})$ would simply be an affine transformation.
We consider quadrilateral element shapes in case of 2D geometries and hexahedral elements in the 3D case. The multi-dimensional shape functions are written as the product of one-dimensional shape functions,~$N_i(\boldsymbol{\xi}) = N_{i_1 ... i_d}(\boldsymbol{\xi}) = l_{i_1}(\xi_1)\cdot ... \cdot l_{i_d}(\xi_d)$. Since we use a nodal DG approach in the present work, the functions~$l(\xi)$ are Lagrange polynomials, i.e., the solution coefficients~$u_i^e$ represent the numerical solution at the ~$N_{\mathrm{DoFs},e}=\left(k+1\right)^d$ interpolation points of the Lagrange polynomials of degree~$k$ in~$d$ dimensions. The approximation quality of the numerical scheme -- also denoted as spatial resolution in the following -- mainly depends on two parameters, the element size~$h$ and the polynomial degree~$k$ of the shape functions.

The weak discontinuous Galerkin formulation can then be derived in two steps: Firstly, the residual of the Navier--Stokes equations is required to be orthogonal to all weighting functions, which are chosen equal to the space of solution functions. Secondly, integration by parts of spatial derivative operators is performed and physical flux functions are replaced by numerical fluxes. By the use of numerical fluxes, information from adjacent elements is combined in order to enforce continuity of the solution in a weak sense. We do not elaborate the weak DG formulation in detail in the present work but refer to~\cite{Hesthaven2007} for a basic description of the general methodology. In the context of the incompressible Navier--Stokes equations, discontinuous Galerkin discretizations have been proposed, e.g., in~\cite{Hesthaven2007,Girault2005b,Shahbazi2007,Bassi2007,Cockburn2009,Ferrer2011,Nguyen2011,Lehrenfeld2016,Fehn2017,Fehn2018a}. The incompressible Navier--Stokes high-order DG solver used in the present study has been developed in~\cite{Fehn2017,Fehn2018a} where a detailed description of weak forms, numerical fluxes, and the imposition of boundary conditions can be found. The weak formulation of the Navier--Stokes equations with the pressure-correction scheme used for discretization in time can be summarized as follows:
Find~$\hat{\bm{u}}_h,\bm{u}_h^{n+1}\in\mathcal{V}^u_h$ and~$\phi_h^{n+1},p_h^{n+1}\in\mathcal{V}^p_h$  such that for all~$\bm{v}_h \in \mathcal{V}^{u}_{h,e}$,~$q_h \in \mathcal{V}^{p}_{h,e}$ and for all elements~$e=1,...,N_{\text{el}}$
\begin{align}
\begin{split}
\intele{\bm{v}_h}{\frac{\frac{3}{2}\hat{\bm{u}}_h-2\bm{u}^{n}_h+\frac{1}{2}\bm{u}^{n-1}_h}{\Delta t}}
+ c^e_h\left(\bm{v}_h,\hat{\bm{u}}_h\right) + v^e_h\left(\bm{v}_h,\hat{\bm{u}}_h\right)
=& - g^e_h\left(\bm{v}_h,p^{n}_h\right)  + \intele{\bm{v}_h}{\bm{f}(t_{n+1})}  \; ,
\end{split} \label{PressureCorrection_MomentumImplicit_Nonlinear_WeakForm}\\
l_{h}^{e}\left(q_h,\phi_h^{n+1}\right) =&
- \frac{3}{2 \Delta t} d_{h}^{e}\left(q_h,\hat{\bm{u}}_h\right)\; ,
\label{PressureCorrection_PressureStep_WeakForm}\\
\begin{split}
\intele{q_h}{p_h^{n+1}} =& \intele{q_h}{\phi_h^{n+1} + p_h^{n}} - \nu\; d_{h}^{e}\left(q_h,\hat{\bm{u}}_h\right) \; ,
\end{split}
\label{PressureCorrection_PressureUpdate_WeakForm}\\
\intele{\bm{v}_h}{\bm{u}_h^{n+1}} + a^e_{\mathrm{D}}(\bm{v}_h,\bm{u}^{n+1}_h) + a^e_{\mathrm{C}}(\bm{v}_h,\bm{u}^{n+1}_h)
=& \intele{\bm{v}_h}{\hat{\bm{u}}_h} -\frac{2 \Delta t}{3} g_h^{e}\left(\bm{v}_h,\phi_h^{n+1}\right)\; .\label{PressureCorrection_Projection_WeakForm}
\end{align}
In the above equations, integrals over an element~$\Omega_e$ (and analogously over the boundary~$\partial \Omega_e$) are abbreviated as~$\intele{v}{u} = \int_{\Omega_e} v \odot u \; \mathrm{d}\Omega$ with the operator~$\odot$ denoting inner products for two scalar, vectorial, or tensorial quantities~$u,v$. Following~\cite{Shahbazi2007}, the local Lax--Friedrichs flux is used as numerical flux function for the nonlinear convective term~$c^e_h\left(\bm{v}_h,\bm{u}_h\right)$. The discretization of the viscous terms~$ v^e_h\left(\bm{v}_h,\bm{u}_h\right)$ as well as the negative Laplace operator~$l_{h}^{e}\left(q_h,p_h\right)$ in the pressure Poisson equation are based on the symmetric interior penalty method~\cite{Arnold2002}. To obtain the weak formulation of the pressure gradient term~$g^e_h\left(\bm{v}_h,p_h\right)$ and the velocity divergence term~$d_{h}^{e}\left(q_h,\bm{u}_h\right)$, integration by parts is performed and central fluxes are used~\cite{Fehn2017}. The weak formulations of the different operators can then be written as~\cite{Fehn2017,Fehn2018a}
\begin{align}
c^e_h\left(\bm{v}_h,\bm{u}_h\right) &= -\intele{\Grad{\bm{v}_h}}{\bm{F}_{\mathrm{c}}(\bm{u}_h)} +
\inteleface{\bm{v}_h}{\left(\avg{\bm{F}_{\mathrm{c}}(\bm{u}_h)} + \max \left(\vert \bm{u}_h^{-} \cdot \bm{n}\vert ,\vert \bm{u}_h^{+} \cdot \bm{n}\vert\right)\jump{\bm{u}_h}  \right) \cdot \bm{n}}\\
g^e_h\left(\bm{v}_h,p_h\right) &= -\intele{\Div{\bm{v}_h}}{p_h}+\inteleface{\bm{v}_h}{ \avg{p_h}\bm{n}}\; ,\\
d^e_h\left(q_h,\bm{u}_h\right) &= -\intele{\Grad{q_h}}{\bm{u}_h}+\inteleface{q_h}{\avg{\bm{u}_h}\cdot\bm{n}}\; ,\\
v_{h}^{e}(\bm{v}_h,\bm{u}_h) &=
 \intele{\Grad{\bm{v}_h}}{\nu \Grad{\bm{u}_h}}
  - \inteleface{\Grad{\bm{v}_h}}{\nu/2\; \jump{\bm{u}_h}}
  - \inteleface{\bm{v}_h}{\nu \avg{\Grad{\bm{u}_h}}\cdot\bm{n}}
  + \inteleface{\bm{v}_h}{\nu\tau \jump{\bm{u}_h}\cdot\bm{n}}\; ,\\
  l_h^e\left(q_h,p_h\right) &= \intele{\Grad{q_h}}{\Grad{p_h}}
-\inteleface{\Grad{q_h}}{1/2\;\jump{p_h}}
- \inteleface{q_h}{\avg{\Grad{p_h}}\cdot\bm{n}}
+ \inteleface{q_h}{\tau\jump{p_h}\cdot\bm{n}}	\; .
\end{align}
Here,~$\avg{u} = (u^- + u^+)/2$ is the average operator and~$ \jump{u} = u^- \otimes \bm{n}^- + u^+ \otimes \bm{n}^+$ the jump operator with~$(\cdot)^-$ denoting interior information,~$(\cdot)^+$ denoting exterior information from the neighboring element, and~$\bm{n}$ the outward pointing unit normal vector. A definition of the interior penalty parameter~$\tau$ can be found in~\cite{Fehn2017}. To ensure inf--sup stability, a mixed-order approach with polynomial degree~$k_u=k$ for the velocity and~$k_p=k-1$ for the pressure is used~\cite{Fehn2017}. A distinctive feature of the present discretization approach is the use of consistent divergence and continuity penalty terms~\cite{Fehn2018a}
\begin{align}
a^e_{\mathrm{D}}(\bm{v}_h,\bm{u}_h) &= \intele{\Div{\bm{v}_h}}{\tau_{\mathrm{D}}\Div{\bm{u}_h}}\; , \\
a^e_{\mathrm{C}}(\bm{v}_h,\bm{u}_h)&=\intelefaceInterior{\bm{v}_h\cdot \bm{n}}{\avg{\tau_{\mathrm{C},e}}\left(\bm{u}^-_h-\bm{u}^+_h\right)\cdot \bm{n}} \; ,
\end{align}
yielding a discretization approach that is robust in the under-resolved regime by improving mass conservation as well as energy stability. In order to focus on the main aspects of the discretization in this work, we refer to~\cite{Fehn2018a} for a detailed definition of the penalty parameters~$\tau_{\mathrm{D}}$ and~$\tau_{\mathrm{C}}$. These penalty terms might be interpreted as a weak enforcement of exactly divergence-free H(div)-conforming function spaces for which discrete energy stability can be shown. The approach can be considered as an implicit or no-model LES strategy without explicit subgrid-scale modeling terms, i.e., the discretization scheme itself provides sufficient numerical dissipation. This strategy has the advantage that it leads to a generic and parameter-free Navier--Stokes solver that is accurate when solving turbulent flow problems but also reproduces the exact solution when applied to laminar flow problems, see~\cite{Fehn2018a} for a detailed discussion. The present DG solver has been extensively verified in~\cite{Fehn2017} for laminar flow problems and in~\cite{Fehn2018a} for turbulent flow problems. The aim of the present study is a validation of this high-order DG solver for typical medical device flow problems involving laminar, transitional, and turbulent flow problems by the example of the FDA benchmark nozzle problem.

\paragraph{Calculation of integrals and implementation aspects} The weak DG formulation contains volume and surface integrals for the different terms of the Navier--Stokes equations. These integrals are transformed from physical space to reference space where they are calculated numerically by means of Gaussian quadrature. The number of quadrature points is selected so that integrals are evaluated exactly on affine element geometries. More detailed information on the chosen quadrature formulas and the number of quadrature points can be found in~\cite{Fehn2017,Fehn2018a}. To obtain a computationally efficient discontinuous Galerkin solver, a fast evaluation of weak forms is crucial, especially for high polynomial degrees. For the tensor product elements considered here, efficient matrix-free implementations are available based on the sum-factorization technique. The present Navier--Stokes solver makes use of a highly-optimized matrix-free implementation developed in~\cite{Kronbichler2012,Kronbichler2017b} and available in the object-oriented finite element library~\texttt{deal.II}~\cite{dealII90}. This hardware-aware implementation results in excellent performance characteristics where the computational costs per unknown for evaluating weak forms are almost independent of the polynomial degree in the range~$2\leq k \leq 10$. This constitutes a perfect basis for not only achieving improved accuracy of high-order methods but also an improved overall efficiency for high-order methods applied to the solution of turbulent flow problems. A detailed performance evaluation of the high-order discontinuous Galerkin solver used here can be found in~\cite{Fehn2018b} by the example of the three-dimensional Taylor--Green benchmark test case and parallel scalability of the algorithm up to large core counts has been shown in~\cite{Krank2017}. Nonlinear systems of equations are solved by a Newton--Krylov approach and linear(ized) systems of equations by state-of-the-art iterative solution techniques such as CG or GMRES with efficient matrix-free preconditioners. Relative solver tolerances of~$\texttt{reltol}=10^{-3}$ and absolute solver tolerances of~$\texttt{abstol}=10^{-12}$ are used in this work, where relative solver tolerance means that the norm of the residual is reduced by~$\texttt{reltol}$ as compared to an initial guess which is an extrapolation from previous instants of time.

\section{Description of test case and numerical setup}\label{TestCase}
This section describes the FDA benchmark nozzle problem and discusses the chosen numerical setup as a prerequisite for the numerical results shown in Section~\ref{NumericalResults}. We first describe the geometry and the boundary conditions in Section~\ref{GeometryBoundaryConditions}. The range of Reynolds numbers and flow regimes are discussed in Section~\ref{ReynoldsNumbers}. The precursor simulation approach is detailed in Section~\ref{Precursor} and the applied initial conditions are described in Section~\ref{InitialConditions}. The mesh is explained in Section~\ref{Mesh} and the calculation of the time step size in Section~\ref{TimeStepSize}. Finally, the statistical sampling of results is summarized in Section~\ref{TimeIntervalAndSampling}.

\subsection{Geometry and boundary conditions}\label{GeometryBoundaryConditions}
\begin{figure}[!ht]
\centering
  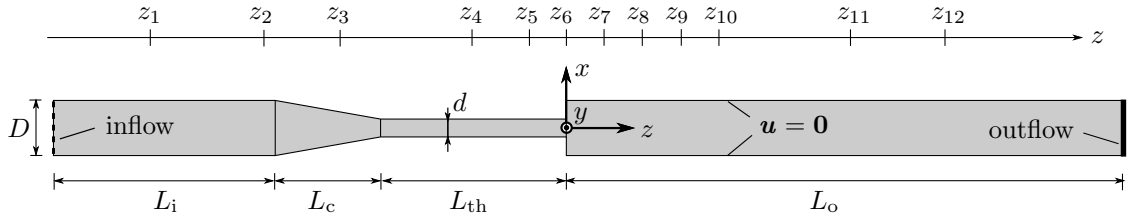
\caption{Geometry of FDA benchmark nozzle model for the sudden expansion test case (flow direction from left to right) and imposed boundary conditions. To assess the accuracy of the numerical approach, results are extracted at various~$z$-locations:~$z_1=-0.088\unit{m}$,~$z_2=-0.064\unit{m}$,~$z_3=-0.048\unit{m}$,~$z_4=-0.02\unit{m}$,~$z_5=-0.008\unit{m}$,~$z_6=0.0\unit{m}$,~$z_7=0.008\unit{m}$,~$z_8=0.016\unit{m}$,~$z_9=0.024\unit{m}$,~$z_{10}=0.032\unit{m}$,~$z_{11}=0.06\unit{m}$, and~$z_{12}=0.08\unit{m}$.}
\label{fig:Geometry}
\end{figure}

The geometry of the FDA benchmark nozzle model with sudden expansion is illustrated in Figure~\ref{fig:Geometry}. The nozzle geometry has been designed so that it contains characteristic features of medical device flow problems such as accelerating and decelerating flows by gradual or sudden changes in the cross-section area, separating flows, as well as transitional and turbulent flows~\cite{Malinauskas2017,Stewart2012}. Principally, the FDA benchmark nozzle model can be used in a bidirectional way with either a sudden expansion or a sudden contraction depending on the flow direction. In the present work, we entirely focus on the sudden expansion test case with flow from left to right in Figure~\ref{fig:Geometry}.
At the wall boundaries, no-slip boundary conditions are prescribed by setting the velocity to zero,~$\bm{u}=\bm{0}$. An inflow boundary condition is prescribed at the left boundary where a precursor simulation approach is used in the present work to generate the inflow velocity profile. A detailed description of the precursor simulation approach is given in Section~\ref{Precursor}. At the right boundary, an outflow boundary condition according to~\cite{Gravemeier2012} is prescribed to obtain a stable discretization approach in case of a turbulent flow in the outflow section. This boundary condition is essential if backflow occurs at the outflow boundary in which case standard outflow boundary conditions prescribing zero traction might become unstable. Between the inlet and throat sections, the flow is accelerated in a conical section by a gradual decrease in the cross-section area. Behind the sudden expansion at streamwise location~$z=0$, the flow enters the outflow section in form of a jet. The main physical dimensions describing the geometry are the throat diameter~$d=0.004 \unit{m}$ and the outer diameter~$D=3d$ of the inlet and outlet sections. The length of the cone and throat sections are defined as~$L_{\mathrm{c}} = \frac{D-d}{2\arctan(\alpha/2)}$ and~$L_{\mathrm{th}}=10 d$, respectively, where~$\alpha=20^{\circ}=\pi/9\ \unit{rad}$ denotes the cone angle. The length~$L_{\mathrm{i}}$ of the inflow section and the length~$L_{\mathrm{o}}$ of the outflow section are not specified by the benchmark. For the inflow section, a value of~$L_{\mathrm{i}} = 4 D$ is used which is sufficiently long since we use a precursor simulation approach to generate the inflow boundary condition and therefore do not need excessively long inflow sections to obtain a developed flow. For the outflow section, a length of~$L_{\mathrm{o}} = 10 D$ is used in the present work which is in a similar range as in previous LES studies~\cite{Delorme2013,Passerini2013,Janiga2014,Zmijanovic2017}.

\subsection{Reynolds numbers and flow regimes}\label{ReynoldsNumbers}
The throat Reynolds number~$\mathrm{Re}_{\mathrm{th}}$, the inlet Reynolds number~$\mathrm{Re}_{\mathrm{i}}$, and the flow rate~$Q$ are defined as
\begin{align*}
\mathrm{Re}_{\mathrm{th}}=\frac{\bar{u}_{\mathrm{th}}d}{\nu} \ , \ \mathrm{Re}_{\mathrm{i}}=\frac{\bar{u}_{\mathrm{i}}D}{\nu} \ , \ Q = \bar{u}_{\mathrm{i}} \frac{D^2 \pi}{4} \ ,
\end{align*}
where~$\bar{u}_{\mathrm{i}}$ and~$\bar{u}_{\mathrm{th}}$ are mean velocities averaged over the inflow and the throat cross-section areas, respectively. Given the throat Reynolds number~$\mathrm{Re}_{\mathrm{th}}$, the inlet Reynolds number~$\mathrm{Re}_{\mathrm{i}}$ and the flow rate~$Q$ can be expressed as follows
\begin{align*}
\mathrm{Re}_{\mathrm{i}}=\mathrm{Re}_{\mathrm{th}} \frac{d}{D} \ , \ Q = \mathrm{Re}_{\mathrm{th}} \frac{d \pi \nu}{4} \ .
\end{align*}
The kinematic viscosity is given as~$\nu=3.31 \cdot 10^{-6} \frac{\unit{m}^2}{\unit{s}}$ in~\cite{Stewart2012}, so that the mean velocities leading to a desired flow rate or Reynolds number can be calculated from the above equations. In Table~\ref{tab:ReynolsNumbersAndFlowRates} we list the throat and inflow Reynolds numbers, the flow rate, and the expected flow regimes (based on experimental results) in the inflow and outflows sections for the different test cases considered in this work.
\begin{table}
\caption{Throat Reynolds number~$\mathrm{Re}_{\mathrm{th}}$, inlet Reynolds number~$\mathrm{Re}_{\mathrm{i}}$, and flow rate~$Q$ for the test cases of the FDA nozzle benchmark. The different test cases are characterized in terms of the expected flow regimes (based on experimental results) in the inflow and outflow sections of the nozzle domain.}
\label{tab:ReynolsNumbersAndFlowRates}
\renewcommand{\arraystretch}{1.1}
\begin{center}
\begin{small}
\begin{tabular}{lllll}
\hline
$\mathrm{Re}_{\mathrm{th}}$ & $\mathrm{Re}_{\mathrm{i}}$ & $Q$ in~$\frac{\unit{m}^3}{\unit{s}}$ & flow regime at inflow & flow regime at outflow\\
\hline
500  & 167  & $5.21 \cdot 10^{-6}$ & laminar & laminar\\
2000 & 667  & $2.08 \cdot 10^{-5}$ & laminar & transitional\\
3500 & 1167 & $3.64 \cdot 10^{-5}$ & laminar & turbulent\\
5000 & 1667 & $5.21 \cdot 10^{-5}$ & laminar & turbulent\\
6500 & 2167 & $6.77 \cdot 10^{-5}$ & transitional & turbulent\\
%8000 & 2667 & $8.32 \cdot 10^{-5}$ & turbulent & turbulent\\
\hline
\end{tabular}
\end{small}
\end{center}
\renewcommand{\arraystretch}{1}
\end{table}

\subsection{Precursor simulation approach}\label{Precursor}

\begin{figure}[!ht]
\centering
  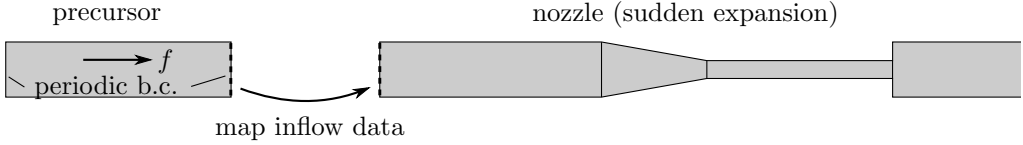
\caption{Illustration of precursor simulation approach used to generate inflow boundary conditions for the actual nozzle domain. In the precursor domain, periodic boundary conditions are prescribed in streamwise direction and the flow is driven by a body force~$f$ in streamwise direction. In each time step, the velocity field is mapped from the precursor to the inflow boundary of the nozzle domain to obtain the (turbulent) inflow boundary condition for the actual computational domain.}
\label{fig:Precursor}
\end{figure}

As already mentioned in Section~\ref{Motivation}, synthetic turbulence generators as well as precursor simulations can be used to prescribe turbulent inflow boundary conditions. As shown in the previous section, the considered Reynolds numbers span the range from laminar to turbulent flows at the inflow boundary. Instead of prescribing a parabolic velocity profile at the inflow boundary or applying other synthetic turbulence inflow boundary conditions, a precursor simulation strategy is used in this work. This approach has the advantage that it does not introduce or require knowledge about the flow regime (laminar, transitional, turbulent), i.e., it is a generic strategy applicable to the full range of Reynolds numbers and flow regimes. We simulate a pipe flow on a precursor domain (diameter~$D$ and length~$L_{\mathrm{p}}=4 D$) with periodic boundary conditions, which is driven by a spatially constant body force that is dynamically adjusted during the simulation such that the desired flow rate is obtained. Then, the velocity field is mapped from the outflow boundary of the precursor domain (but any other cross-section could be used as well) to the inflow boundary of the actual nozzle domain in each time step as illustrated in Figure~\ref{fig:Precursor}. It is often argued that this approach has the disadvantage that it requires additional computational costs. Note, however, that long inflow domains often required to obtain a developed turbulent flow can be avoided by such a technique. Hence, it is unclear whether this approach increases computational costs. Apart from that, investing additional effort into a percursor simulation appears to be reasonable in view of the potentially improved accuracy that can be achieved with such an approach, i.e., a more expensive but physically correct simulation might be more efficient overall than a cheaper but less accurate simulation. For example, when prescribing wrong inflow boundary conditions the results might not even converge to the exact solution no matter how fine the mesh is. We comment on this aspect in more detail in Section~\ref{NumericalResults}. By applying the strategy detailed above, we obtain a generic and parameter-free incompressible flow solver.

\paragraph{Flow rate controller} To ensure that the actual mass flow rate through the nozzle domain~$Q_{\mathrm{m}}$ follows the desired flow rate~$Q$, we use a flow rate controller for the precursor domain that dynamically adjusts the body force~$\bm{f} = (0,0,f)^T$ acting in downstream direction~$z$ in the precursor domain, see Figure~\ref{fig:Precursor}. The body force~$f$ is given as
\begin{align*}
f = f_0 + \int_{t_0}^{t} K \left(Q - Q_{\mathrm{m}}(t)\right) \mathrm{d}t \approx f_0 + \sum_{i=0}^{N_i} K \left(Q - Q_{\mathrm{m}}(t_i)\right) \Delta t \; ,
\end{align*}
i.e., the body force is increased if the measured flow rate is lower than the target flow rate~$Q$ and vice versa. In the above equation, the time step number is denoted by~$N_i$. The initial guess~$f_0$ is chosen as~$f_0 = 8 \nu \bar{u}_{\mathrm{i}} / (D/2)^2$, which is the body force that would result in the desired flow rate under the assumption of a parabolic velocity profile in radial direction. Note that this choice is in agreement with the chosen initial conditions for the velocity field. The proportionality constant~$K$ with physical unit~$\left[ K\right] = 1/\left(\mathrm{m}^2 \mathrm{s}^2\right)$ is derived by means of dimensional analysis and is expressed as a function of the mean velocity~$\bar{u}_{\mathrm{i}}$ and the diameter~$D$ to obtain
%For completeness, the other quantities have physical units of~$\left[f\right] = \mathrm{m}/\mathrm{s}^2$,~$\left[ Q\right] = \mathrm{m}^3/\mathrm{s}$, and~$\left[ \Delta t\right]= \mathrm{s}$.
\begin{align*}
K = C \frac{\bar{u}_{\mathrm{i}}^2}{D^4} \; , \; \left[ K \right] = \frac{\left(\mathrm{m}/\mathrm{s}\right)^2}{\mathrm{m}^4} = \frac{1}{\mathrm{m}^2 \mathrm{s}^2} \; .
\end{align*}
This definition has the advantage that one does not have to readjust this constant for different flow parameters such as the Reynolds number. For the remaining constant~$C$, we found that a value of~$C=1$ ensures an accurate flow rate and is therefore used for all simulations performed in this work. After each time step, the measured flow rate~$Q_\mathrm{m}$ is calculated as the volume-averaged velocity in~$z$-direction in the precursor domain multiplied by the inflow area~$A_{\mathrm{i}}$.

\subsection{Initial conditions}\label{InitialConditions}

The~$z$-component of the velocity field is initialized with a parabolic velocity profile in both the precursor domain as well as the actual nozzle domain
\begin{align*}
u_z(r,z,t=0) = 2 \bar{u}(z) \left(1-\left(\frac{r}{r(z)}\right)^2 \right) \; ,
\end{align*}
where~$\bar{u}(z) = Q/A(z)$ is the mean velocity in streamwise direction at location~$z$ averaged over the cross-section area~$A(z)=r^2(z) \pi$, and~$r = \sqrt{x^2+y^2} \leq r(z)$ the distance of a point from the~$z$-axis. The radius function~$r(z)$ defines the radial location of the no-slip walls as a function of~$z$. In the precursor domain only,  we add random perturbations to the~$z$-component of the velocity field as well as larger vortices prescribed via sine functions with a wavelength of four full periods in circumferential direction as well as in~$z$-direction in order to initiate a turbulent flow. Note, however, that this approach is still generic in the sense that the initially unsteady/turbulent flow will return to a laminar behavior in case that the Reynolds number is not sufficiently large to sustain a turbulent flow. On the other hand, the numerical solution might have difficulties in becoming turbulent if no imperfections are present and if the flow is initialized with a symmetric analytical solution. The other velocity components in~$x$ and~$y$-direction are initialized with zero in both domains. Similarly, the pressure field is~$p(\bm{x}, t_0) = 0$ in both the precursor domain and the actual nozzle domain.

\subsection{Mesh}\label{Mesh}
\begin{figure}[!ht]
 \centering
 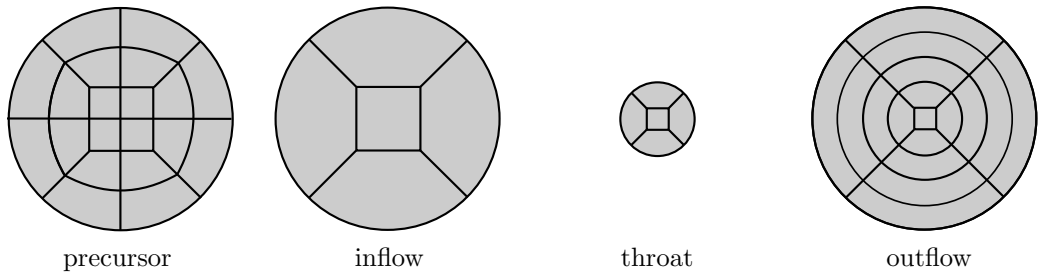
\caption{Visualization of mesh for different cross-sections in the precursor, inflow, throat, and outflow sections. The coarsest mesh with refinement level~$l=0$ is shown here and finer meshes are obtained by uniform mesh refinements with level~$l>0$.}
\label{fig:Mesh}
\end{figure}

A structured mesh composed of non-overlapping and conforming hexahedral elements is used. Basically, the mesh resolution can be described as a function of a characteristic grid size~$h$ and the polynomial degree~$k$ of the shape functions. Alternatively, given a fixed coarse mesh, the mesh resolution can be characterized by the mesh refinement level~$l$ instead of the grid size~$h$. We explicitly construct a mesh for the coarsest refinement level~$l=0$, which is shown in Figure~\ref{fig:Mesh}, and obtain finer meshes by uniformly refining the coarsest mesh~$l$ times. As a result, the overall number of unknowns increases by a factor of~$2^d=8$ in 3D when increasing the mesh refinement level. Note, however, that the overall number of unknowns can be increased more gradually by increasing the polynomial degree~$k$ which can be seen as an advantage of the high-order DG discretization used here, offering flexibility regarding the selection of the polynomial degree~$k$ and hence of the overall problem size.

The coarsest mesh is illustrated in Figure~\ref{fig:Mesh} for different cross-sections in the precursor domain as well as in the inflow, throat, and outflow parts of the nozzle domain. For an accurate representation of cylindrical and conical boundaries, an isoparametric approach is used by analytically prescribing the curved boundaries via volume manifold descriptions and using high-order mappings of the same polynomial degree~$k$ for the mapping from reference space to physical space. In the cone section, the mesh in the~$x$-$y$-plane is stretched linearly as a function of~$z$ to continuously decrease the diameter from~$D$ to~$d$. On the coarsest mesh ($l=0$), the mesh is already refined once in the precursor domain as compared to the inflow section of the nozzle domain. A finer mesh is chosen for the precursor domain in order to ensure an accurate velocity field at the inflow boundary. In~$z$-direction, the elements are distributed uniformly both in the precursor domain and in all parts of the nozzle domain. The number of elements in~$z$-direction for mesh refinement level~$l=0$ is~$N_{\mathrm{ele},\mathrm{p},z}(l=0) = 16$ in the precursor,~$N_{\mathrm{ele},\mathrm{i},z}(l=0) = 8$ in the inflow section,~$N_{\mathrm{ele},\mathrm{c},z}(l=0) = 4$ in the cone section,~$N_{\mathrm{ele},\mathrm{th},z}(l=0) = 8$ in the throat section, and~$N_{\mathrm{ele},\mathrm{o},z}(l=0) = 20$ in the outflow section. Consequently, the coarsest mesh consists of~$N_{\mathrm{ele},\mathrm{p}}(l=0) = 320$ for the precursor domain and~$N_{\mathrm{ele},\mathrm{n}}(l=0) = 440$ for the nozzle domain. On finer meshes the number of elements is~$N_{\mathrm{ele}}(l) = N_{\mathrm{ele}}(l=0) \left(2^d\right)^l$, and the overall number of unknowns is~$N_{\mathrm{DoFs}} = N_{\mathrm{ele}}(l) \left( d (k_u+1)^d + (k_p+1)^d \right)$. The overall number of unknowns both in the precursor domain and in the nozzle domain is listed in Table~\ref{tab:MeshesAndDoFs} for various refinement levels~$l$ and polynomial degrees~$k$.

\begin{table}
\caption{Number of degrees of freedom~$N_{\mathrm{DoFs}}$ for both velocity and pressure unknowns for different refinement levels~$l$ and polynomial degrees~$k$. The spatial resolutions analyzed in Section~\ref{NumericalResults} for each throat Reynolds number are also listed.}
\label{tab:MeshesAndDoFs}
\renewcommand{\arraystretch}{1.1}
\begin{center}
\begin{scriptsize}
\begin{tabular}{cccccccccc}
\hline
& & \multicolumn{2}{c}{$N_{\mathrm{DoFs}}$ (velocity+pressure)} & &\multicolumn{5}{c}{Throat Reynolds number~$\mathrm{Re}_{\mathrm{th}}$}\\
\cline{3-4} \cline{6-10}
$l$ & $k$ & precursor & nozzle &  & 500 & 2000 & 3500 & 5000 & 6500\\
\hline
0 & 2 & $2.9\cdot 10^4$ & $3.9\cdot 10^4$ & & \Checkmark & 			  & & & \\
  & 3 & $7.0\cdot 10^4$ & $9.6\cdot 10^4$ & & \Checkmark & 			  & & & \\
  & 4 & $1.4\cdot 10^5$ & $1.9\cdot 10^5$ & & \Checkmark & 			  & & & \\
  & 5 & $2.5\cdot 10^5$ & $3.4\cdot 10^5$ & & \Checkmark & \Checkmark & & & \\
%  & 6 & $4.0\cdot 10^5$ & $5.5\cdot 10^5$ & & & & & & \\
%  & 7 & $6.0\cdot 10^5$ & $8.3\cdot 10^5$ & & & & & & \\
\hline
%1 & 2 & $2.3\cdot 10^5$ & $3.1\cdot 10^5$ & & & & & & \\
1 & 3 & $5.6\cdot 10^5$ & $7.7\cdot 10^5$ & & & \Checkmark & \Checkmark & 			 & 			  \\
  & 4 & $1.1\cdot 10^6$ & $1.6\cdot 10^6$ & & & \Checkmark & \Checkmark & 			 & 			  \\
  & 5 & $2.0\cdot 10^6$ & $2.7\cdot 10^6$ & & & \Checkmark & \Checkmark & \Checkmark & 		      \\
  & 6 & $3.2\cdot 10^6$ & $4.4\cdot 10^6$ & & & 		   & 			& \Checkmark & \Checkmark \\
%  & 7 & $4.8\cdot 10^6$ & $6.6\cdot 10^6$ & & & & & & \\
\hline
%2 & 2 & $1.8\cdot 10^6$ & $2.5\cdot 10^6$ & & & & & & \\
2 & 3 & $4.5\cdot 10^6$ & $6.2\cdot 10^6$ & & & \Checkmark & \Checkmark & \Checkmark & \Checkmark \\
  & 4 & $9.0\cdot 10^6$ & $1.2\cdot 10^7$ & & & 		   & \Checkmark & \Checkmark & \Checkmark \\
  & 5 & $1.6\cdot 10^7$ & $2.2\cdot 10^7$ & & & 		   & 			& \Checkmark & \Checkmark \\
%  & 6 & $2.6\cdot 10^7$ & $3.5\cdot 10^7$ & & & & & & \\
%  & 7 & $3.9\cdot 10^7$ & $5.3\cdot 10^7$ & & & & & & \\
\hline
3  & 3 & $3.6\cdot 10^7$ & $4.9\cdot 10^7$ & & & 		   & 			& 			 & \Checkmark \\
\hline
\end{tabular}
\end{scriptsize}
\end{center}
\renewcommand{\arraystretch}{1}
\end{table}

\subsection{Time step size}\label{TimeStepSize}
For turbulent flow problems with high spatial resolution requirements one typically observes that the CFL condition is restrictive and that larger time step sizes would be desirable from the point of view of computational costs without impacting the accuracy. For this reason, we use an implicit formulation of the convective term in this work. The time step size is calculated using the CFL criterion~\cite{Fehn2018a,Fehn2018b}
\begin{align*}
\Delta t = \frac{\mathrm{Cr}}{k_u^{1.5}}\frac{h_{\mathrm{min}}}{\Vert \bm {u} \Vert_{\mathrm{max}}} \; .\label{CFL_Condition}
\end{align*}
Since we use a fully-implicit formulation, a Courant number larger than the critical value corresponding to an explicit treatment of the convective term,~$\mathrm{Cr}>\mathrm{Cr}_{\mathrm{crit}}$, can be used in order to reduce computational costs as compared to a semi-implicit formulation. For all numerical experiments performed in this work, a value of~$\mathrm{Cr}=4$ is used. In the above equation, ~$\Vert \bm {u} \Vert_{\mathrm{max}}$ is an estimation of the maximum velocity and is calculated as~$\Vert \bm {u} \Vert_{\mathrm{max}}=2 \bar{u}_{\mathrm{th}}$. The characteristic element length scale~$h_{\mathrm{min}}$ is calculated as the minimum distance between two vertices of the mesh. Since the precursor domain and the nozzle domain are advanced in time simultaneously, the same time step size is used for both simulations by taking the minimum time step sizes of the two domains.

\subsection{Simulated time interval and sampling of statistical results}\label{TimeIntervalAndSampling}
% NOTE: use a longer precursor of length~$L_{\mathrm{p}}=4 D$, since we observed that unsteady/turbulent structures had a length as long as the precursor domain is case of~$L_{\mathrm{p}}=2 D$, i.e., the length of the precursor domain might impact the results although the results should be independent of length of the precursor domain.

The simulated time interval is chosen as a multiple of a characteristic flow-through time~$t_{\mathrm{c}} = L/u$ expressed as a function of a characteristic length scale~$L$ and a characteristic velocity~$u$. Since the length of the inflow and outflow sections and hence the overall length of the nozzle domain are not defined by the benchmark, we choose the length~$L_{\mathrm{th}}$ of the throat section as characteristic length scale and the mean velocity~$\bar{u}_{\mathrm{th}}$ as reference velocity, yielding the flow-through time~$t_{\mathrm{c}}$
\begin{align*}
t_{\mathrm{c}} = \frac{L_{\mathrm{th}}}{\bar{u}_{\mathrm{th}}} \ .
\end{align*}
Before starting the simulation on the actual nozzle domain, the flow is simulated in the precursor domain over the time interval~$-500 t_{\mathrm{c}} \leq t \leq 0$ to make sure that a developed and statistically steady state flow is reached in the precursor domain. Subsequently, both precursor and nozzle are simulated simultaneously over the time interval~$0 \leq t \leq 250 t_{\mathrm{c}}$, where the statistical sampling of the results to obtain time-averaged quantities is performed over the time interval~$50 t_{\mathrm{c}} \leq t \leq  250 t_{\mathrm{c}}$. With this setup, the statistical errors -- which are largest in the outflow section where the effective number of flow-through times is considerably smaller due to the reduced streamwise velocity related to the increase in the cross-section area -- can be expected to be small. We found that a rather short time interval of~$ 0 \leq t \leq 50 t_{\mathrm{c}}$ is sufficient to obtain a developed flow in the nozzle domain since the inflow boundary condition has already reached a statistically steady state at~$t=0$. To evaluate the accuracy of the numerical results, we consider the mean (time-averaged) streamwise velocity~$\left< u_z (r=0,z)\right>$ along the~$z$-axis as well as radial profiles of the streamwise velocity~$\left< u_z(r,z_i)\right>$ at various~$z_i$-locations,~$i=1,...,12$, as illustrated in Figure~\ref{fig:Geometry}. Statistical mean values are calculated by sampling the results every time step. For the radial profiles, the velocity field is additionally averaged in circumferential direction due to the rotational symmetry of the problem. In the following, all profiles are presented in non-dimensional form using~$\bar{u}_{\mathrm{th}}$ as reference velocity for the streamwise velocity along the~$z$-axis,~$\left< u_z (r=0,z)\right>/\bar{u}_{\mathrm{th}}$, and the mean (averaged over cross-section) velocity~$\bar{u}(z_i)$ as introduced in Section~\ref{InitialConditions} for the radial profiles,~$\left< u_z(r,z_i)\right>/\bar{u}(z_i)$.

\begin{remark}
The non-dimensionalization of velocity profiles is used for an intuitive interpretation of velocity profiles shown in the following. Without changes in the cross-section area, one would expect the mean velocity to take a value between~$1$ and~$2$ on the centerline ($r=0$), where a value of~$2$ would correspond to a laminar parabolic profile and a value of~$1$ would be reached in case of a velocity profile that is constant over the whole cross-section. Behind the sudden expansion, the flow forms a jet and does not immediately adapt to the change in the cross-section. However, since the mean velocity~$\bar{u}(z_i)$ is based on the actual cross-section~$A(z_i)$ and the cross-section increases by a factor of~$(D/d)^2=9$ behind the sudden expansion, the non-dimensional velocity may therefore take a maximum value between~$9$ and~$18$ on the centerline, see also the examples in Section~\ref{NumericalResults}. Behind the location of the jet breakdown, the full cross-section is again utilized, and the non-dimensional velocity recovers a maximum value between~$1$ and~$2$.
\end{remark}

\section{Numerical results}\label{NumericalResults}
\begin{figure}[!ht]
 \centering
 \includegraphics[width=1.0\textwidth]{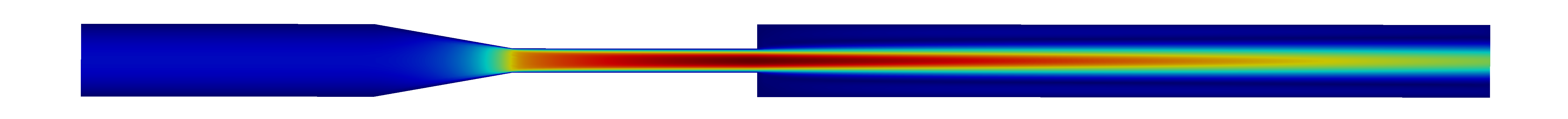}
 \includegraphics[width=1.0\textwidth]{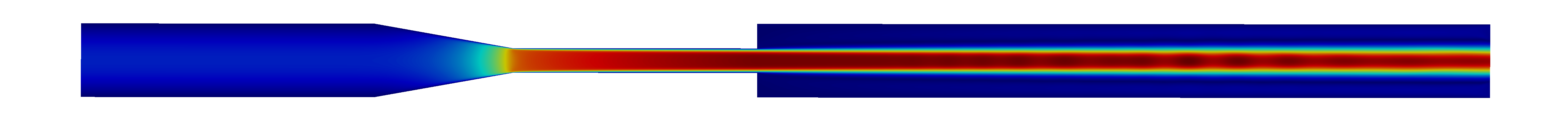}
 \includegraphics[width=1.0\textwidth]{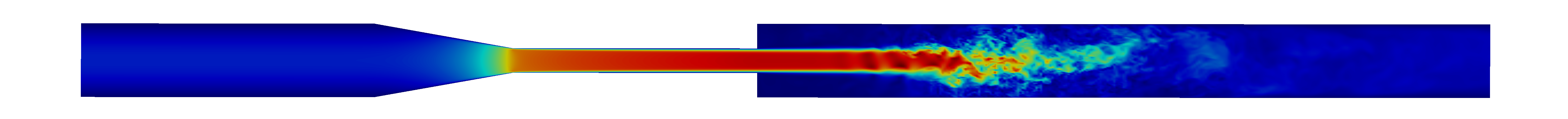}
 \includegraphics[width=1.0\textwidth]{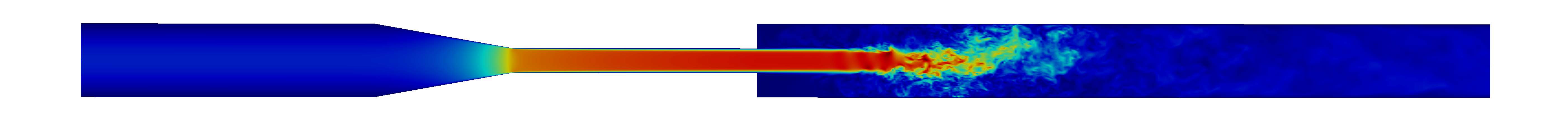}
 \includegraphics[width=1.0\textwidth]{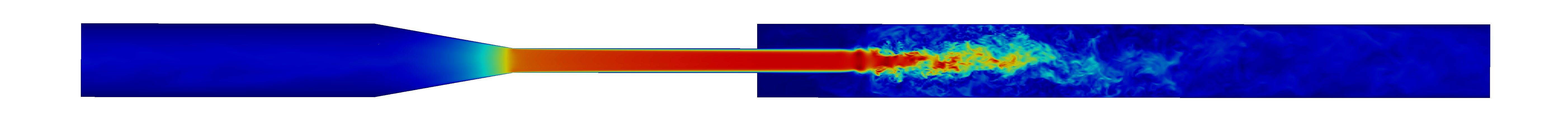}
\caption{Visualization of numerical results for FDA benchmark nozzle model (sudden expansion) for throat Reynolds numbers of~$\mathrm{Re}_{\mathrm{th}}=500, 2000, 3500, 5000, 6500$ (from top to bottom): The figure shows contour plots of the magnitude of instantaneous velocity fields in the~$x$-$y$-plane where blue indicates low velocities and red indicates high velocities (the color bar has been rescaled for each Reynolds number individually).}
\label{fig:velocity_magnitude}
\end{figure}

In this section, numerical results are presented for all Reynolds numbers~$\mathrm{Re}_{\mathrm{th}}=500-6500$ of the FDA benchmark nozzle problem. For each Reynolds number, the numerical results are compared to experimental results~\cite{Hariharan2011} published online~\cite{HariharanExperiments} and a grid convergence study is performed in order to critically assess the accuracy and predicitve capabilities of our approach.
%Additionally, we show results for a larger throat Reynolds number of~$\mathrm{Re}_{\mathrm{th}}=8000$ for which the flow becomes turbulent at the inflow.
The main challenge of this benchmark lies in the correct prediction of the jet breakdown location in axial direction~\cite{Passerini2013,Zmijanovic2017}. Previous LES studies revealed a high sensitivity of the jet breakdown location with respect to parameters of the numerical solution approach~\cite{Zmijanovic2017}, highlighting the necessity to explicitly consider results for a series of grid resolutions in order to substantiate the reliability of the numerical solution. For Reynolds numbers of~$\mathrm{Re}_{\mathrm{th}}=3500, 5000, 6500$, we also show results obtained with a laminar parabolic profile at the inflow boundary as used in previous LES studies instead of the precursor simulation approach in order to illustrate the delicate aspect of inflow boundary conditions and to highlight the role of the precursor simulation approach. In Figure~\ref{fig:velocity_magnitude}, the flow field is visualized for the different Reynolds numbers to give a first qualitative impression of the flow. A quantitative discussion of the results is subject of the subsequent sections.

\subsection{Reynolds number~$\mathrm{Re}_{\mathrm{th}}=500$}
\begin{figure}[!ht]
 \centering
 \subfigure[Profile of mean streamwise velocity along centerline.]{
	\includegraphics[width=0.8\textwidth]{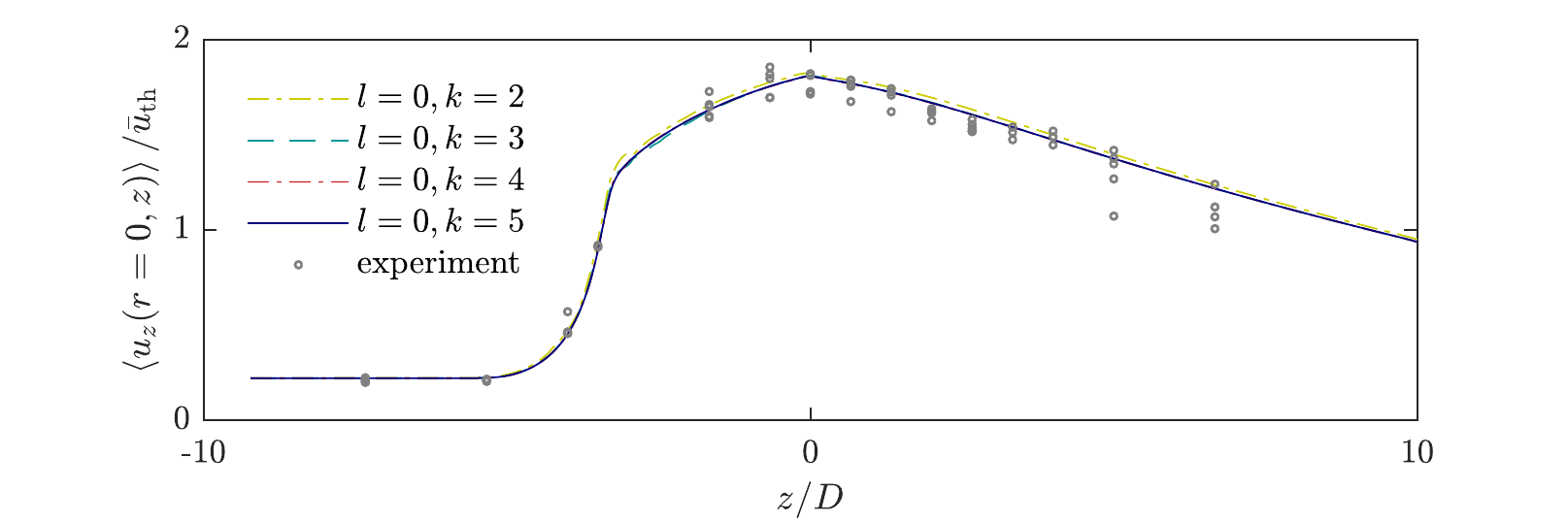}}
 \subfigure[Radial profiles of mean streamwise velocity at various locations~$z_i$,~$i=1,...,12$.]{
	\includegraphics[width=0.8\textwidth]{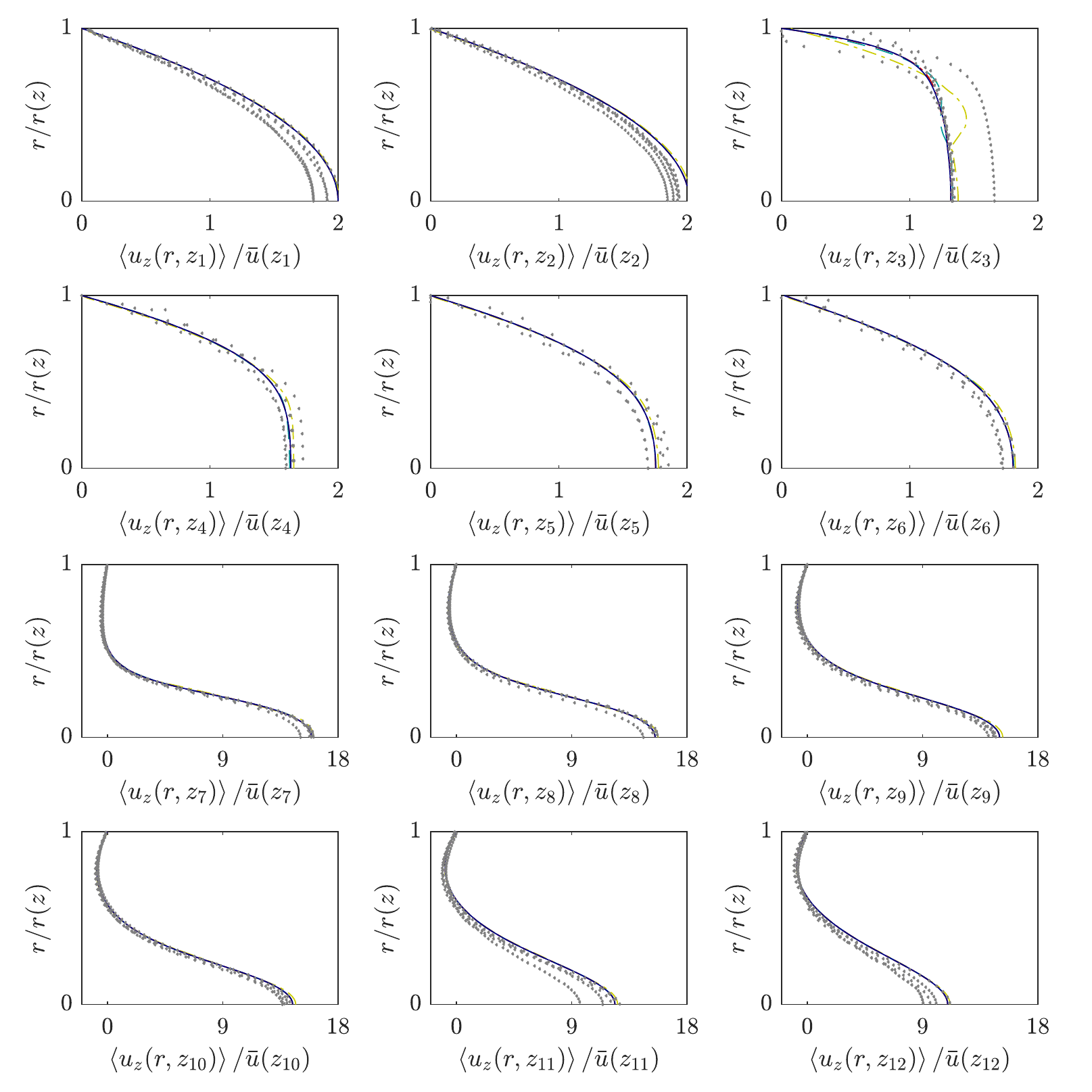}}
\caption{Numerical results for FDA benchmark nozzle model (sudden expansion) at~$\mathrm{Re}_{\mathrm{th}}=500$.}
\label{fig:results_Re500}
\end{figure}
The~$\mathrm{Re}_{\mathrm{th}}=500$ test case is the most simple one from a numerical point of view. Experimental results reveal that the flow remains laminar in all parts of the nozzle geometry~\cite{Hariharan2011}. Although the flow regime is known for this Reynolds number and one therefore tends to prescribe a laminar parabolic velocity profile at the inflow, we use the precursor simulation approach also for this Reynolds number to mimic the original benchmark computations that have been performed in a blinded way. In case that the percursor technique is a generic approach, the flow will be laminar in the precursor domain and a physically correct inflow boundary condition will be prescribed. Numerical results for this test case have been presented for example in~\cite{Delorme2013} and a mesh refinement study has been shown in~\cite{Passerini2013} where a parabolic inflow profile is prescribed in both studies, see Table~\ref{tab:PreviousLES}. Numerical results for the mean streamwise velocity along the centerline ($z$-axis) as well as radial profiles at various~$z$-locations obtained with the present solver are shown in Figure~\ref{fig:results_Re500}. The spatial resolutions considered for this mesh refinement study are listed in Table ~\ref{tab:MeshesAndDoFs}. Already for the coarsest resolution with refinement level~$l=0$ and polynomial degree~$k=2$ the results agree very well with the experimental results. Results for the different spatial resolutions only differ in the radial velocity profile at position~$z_3$ located in the cone section, which can be explained by the fact that the mesh consists of only 5 elements in this cross-section and that the flow is accelerated in this part of the nozzle domain. For increasing polynomial degree~$k$ the results converge rapidly and can be considered as grid-independent for~$k=4,5$. An overall excellent agreement with experimental measurements is obtained for this test case. Small deviations from the experimental results indicate that the experimental results do not exactly match the desired flow rate or Reynolds number, which can be seen from the radial velocity profiles at~$z_1,z_2$ with the maximum of the non-dimensional velocity being smaller than~$2$. At location~$z_3$ one experiment shows a significantly larger mean streamwise velocity. Since this would imply a higher flow rate, this is an indication of errors in the measurements. In the outflow section, the mean streamwise velocity along the centerline is close to the maximum values observed in experiments but is within the range of the experimental results and in agreement with numerical results shown in~\cite{Delorme2013,Passerini2013}.

\subsection{Reynolds number~$\mathrm{Re}_{\mathrm{th}}=2000$}
The test case with throat Reynolds number~$\mathrm{Re}_{\mathrm{th}}=2000$ is more challenging due to its transitional character in the outflow section, see Table~\ref{tab:ReynolsNumbersAndFlowRates}. It has been considered in several previous LES studies~\cite{Delorme2013,Passerini2013,Chabannes2017,Nicoud2018}. None of these works explicitly shows results of a grid refinement study even though the authors of~\cite{Delorme2013} mention that a grid independence study has been carried out for the transitional case. Parabolic velocity profiles are prescribed at the inflow boundary in~\cite{Delorme2013,Passerini2013,Chabannes2017}, while turbulent fluctuations are added to the laminar profile in~\cite{Nicoud2018}. Although excellent agreement with experimental results is reported in previous LES studies, the reliability of the results appears to be unclear due to the limited amount of data provided by these LES studies. In fact, the authors of~\cite{Passerini2013} admit that they ``found the results to be very sensitive to mesh size and time step'' and that they ``managed to identify a mesh'' for which good agreement with experimental results has been obtained. Furthermore, it is mentioned that the use of linear instead of quadratic shape functions resulted in a jet breakdown at a~$z$-location significantly larger than in the experiments. In~\cite{Nicoud2018}, excellent agreement with measurements is obtained by prescribing turbulent fluctuations at the inflow boundary but the robustness of this turbulence injection approach is not demonstrated in that work for the transitional case. In contrast, grid-converged results are reported (but not explicitly demonstrated) in~\cite{Delorme2013} without adding fluctuations to the inflow profile. The following investigations give reasoning for why a thorough numerical investigation of this test case is of high importance in our opinion.
\begin{figure}[!ht]
 \centering
 \subfigure[Profile of mean streamwise velocity along centerline.]{
	\includegraphics[width=0.8\textwidth]{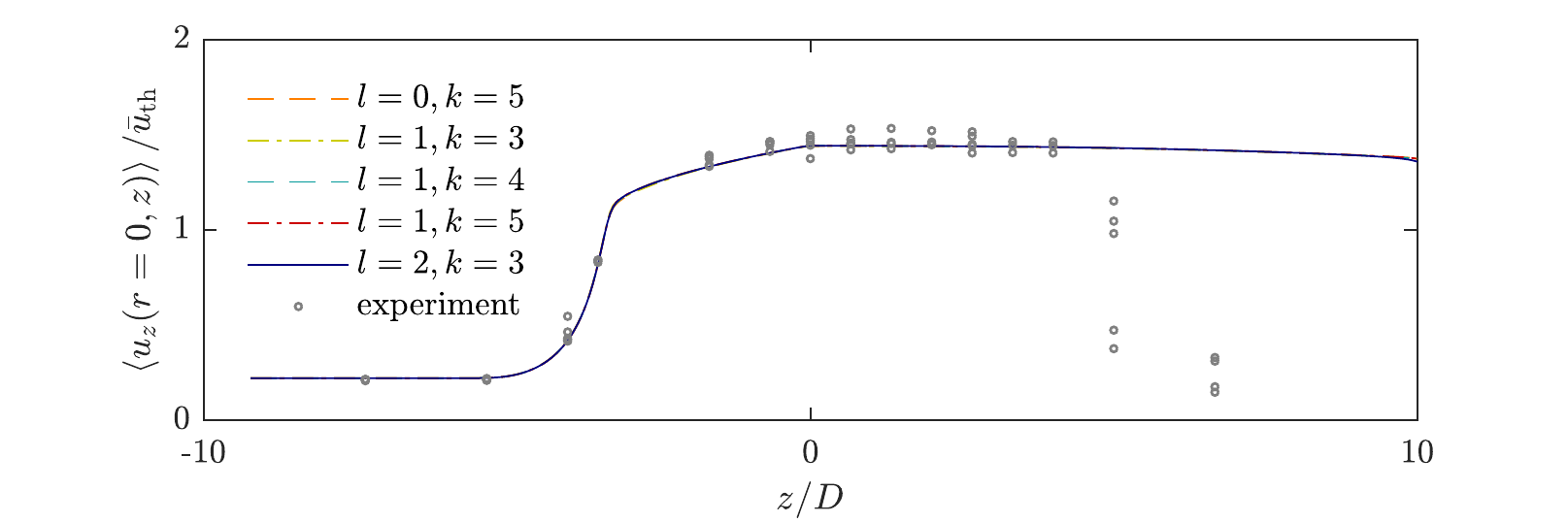}}
 \subfigure[Radial profiles of mean streamwise velocity at various locations~$z_i$,~$i=1,...,12$.]{
	\includegraphics[width=0.8\textwidth]{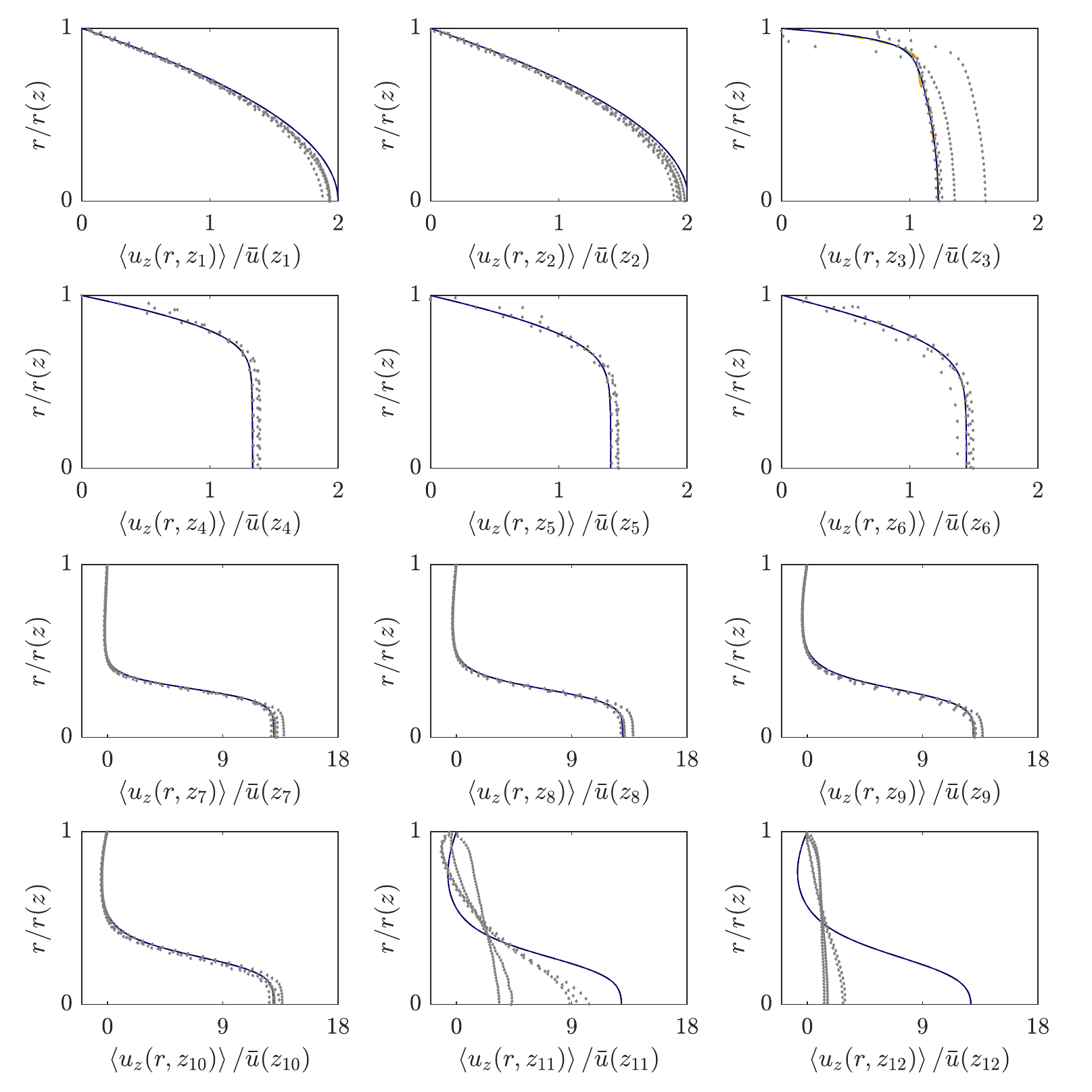}}
\caption{Numerical results for FDA benchmark nozzle model (sudden expansion) at~$\mathrm{Re}_{\mathrm{th}}=2000$.}
\label{fig:results_Re2000}
\end{figure}

In Figure~\ref{fig:results_Re2000}, results obtained with the present high-order DG discretization approach are shown for several spatial resolutions. In the inflow section, cone section, throat section, and close to the sudden expansion in the outflow section, excellent agreement with experimental results is obtained. As for the laminar test case, larger values of the mean streamwise velocity have been measured at location~$z_3$ in some experiments indicating deviations from the target flow rate or Reynolds number. Given the fact that a similar pattern is observed for the laminar test case and the larger Reynolds number test cases shown in the following, it can be conjectured that these discrepancies are related to a systematic error in the measurements. To verify this assumption, the present numerical results could be compared to other numerical studies, but unfortunately none of the previous LES studies has shown results for the radial velocity profile at this specific downstream location~$z_3$. A major difference to the experimental results is that we do not observe a breakdown of the jet in the outflow section which has length~$L_{\mathrm{o}}=10 D$ for the present numerical simulations and the considered spatial resolutions. As a result, large deviations from experimental results can be identified for the radial velocity profiles at streamwise locations~$z_{11}$ and~$z_{12}$. To make sure that this behavior is not related to insufficient spatial resolution, we also simulated the problem on finer meshes with refinement level~$l=2$ and polynomial degrees~$k=4,5$ as well as~$l=3$ and~$k=3$. Here, we observed that at some times the jet breaks down very close to the outflow boundary located at~$z=10D$, but that the jet remains stable over the whole outflow section at other times. While this behavior clearly confirms the transitional character of the flow at this Reynolds number, these simulations do not give an indication that the jet breakdown location will tend towards the experimental results on finer meshes given the fact that the flow is well-resolved for these spatial resolutions and that accurate results are obtained on these meshes for the highest Reynolds number test cases. Interestingly, also the experimental results show large variations at locations~$z_{11}$ and~$z_{12}$ indicating a large sensitivity of the results for this transitional test case. In~\cite{Hariharan2011}, two explanations are provided for the large differences between different experiments conducted in independent laboratories. On the one hand, uncertainties in fluid property measurements or inlet velocity might cause errors (around~$10\%$) in the mass flow rate or the Reynolds number. On the other hand, flow perturbations at the inflow boundary might have a significant impact on the jet breakdown as has been exemplified by the use of fluid tanks of variable size causing significant differences in the fluctuation level and the jet breakdown location.
\begin{figure}[!ht]
 \centering
 \subfigure[Profile of mean streamwise velocity along centerline for different throat Reynolds numbers using the precursor simulation approach.]{
	\includegraphics[width=0.8\textwidth]{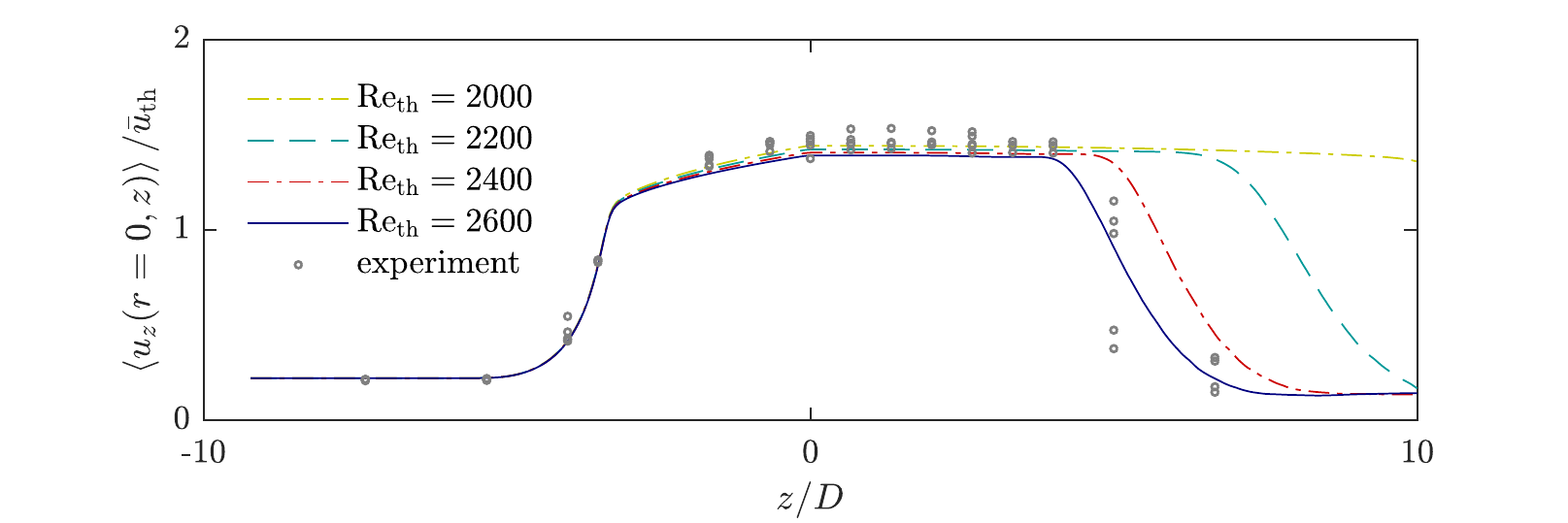}}
 \subfigure[Profile of mean streamwise velocity along centerline prescribing a parabolic inflow profile with random perturbations (specified in~$\%$ of mean velocity at inflow) for a Reynolds number of~$\mathrm{Re}_{\mathrm{th}}=2000$.]{
	\includegraphics[width=0.8\textwidth]{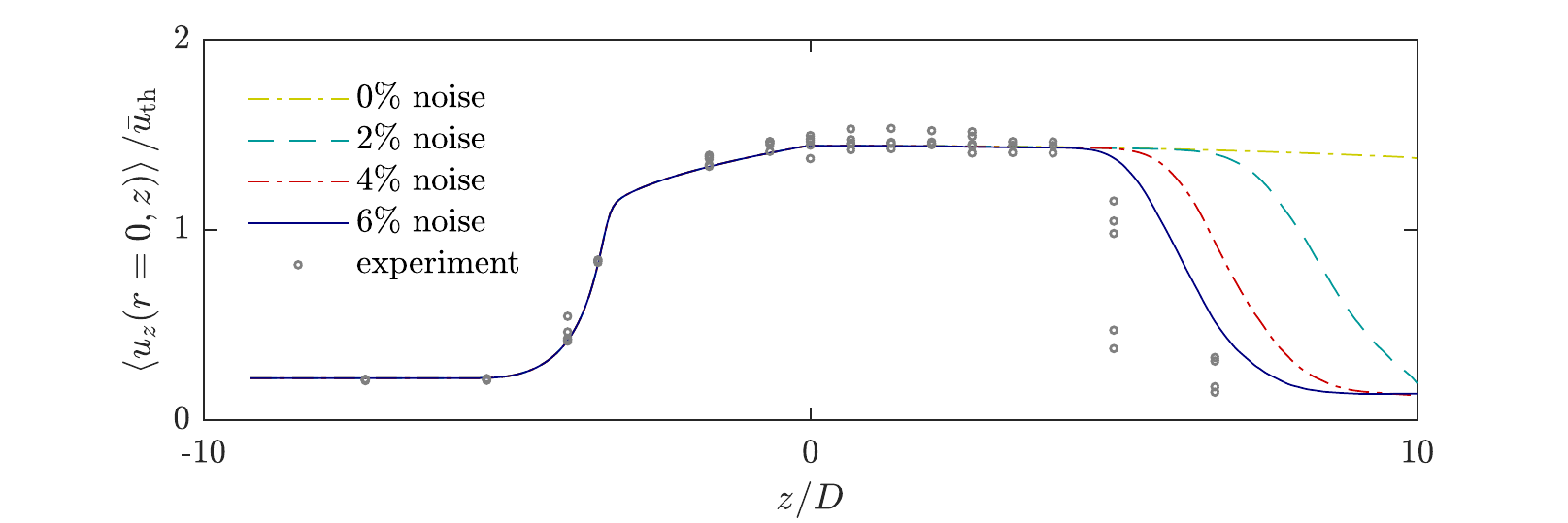}}
\caption{Numerical results for FDA benchmark nozzle model (sudden expansion) at~$\mathrm{Re}_{\mathrm{th}}=2000$: Influence of Reynolds number and inflow fluctuations (random perturbations) on jet breakdown location. The spatial resolution used for these investigations is~$l=2, k=3$.}
\label{fig:results_Re2000_sensitivity}
\end{figure}

To analyze the impact of both aspects, we perform additional simulations using a fixed spatial resolution of~$l=2$ and~$k=3$, i.e., the finest resolution shown in Figure~\ref{fig:results_Re2000}. On the one hand, we simulate the problem for larger Reynolds numbers of~$\mathrm{Re}_{\mathrm{th}}=2200$,~$\mathrm{Re}_{\mathrm{th}}=2400$, and~$\mathrm{Re}_{\mathrm{th}}=2600$ by reducing the viscosity similar to the additional measurements in~\cite{Hariharan2011} in order to test the impact of variations in the Reynolds number. On the other hand, we prescribe a parabolic velocity profile at the inflow with additional small fluctuations similar to~\cite{Zmijanovic2017} in order to simulate disturbances in the flow field at the inflow. For this purpose, we simply prescribe random perturbations (white noise) in streamwise direction with an amplitude of~$0\%$,~$2\%$,~$4\%$, and~$6\%$ of the mean streamwise velocity. The perturbation field at the inflow boundary is recomputed after each time step. Of course, such an approach is insufficient as sophisticated synthetic turbulence generator since it does not introduce physically motivated turbulent structures at the inflow~\cite{Tabor2010}, but this strategy appears to be sufficient for the point that we want to make here. As illustrated in Figure~\ref{fig:results_Re2000_sensitivity}, the jet breakdown location moves closer to the experimental results when increasing the Reynolds number or when increasing the level of disturbances at the inflow. These results are consistent in the sense that the jet breakdown location moves more and more towards the sudden expansion for larger Reynolds numbers of~$\mathrm{Re}_{\mathrm{th}}=3500-6500$ investigated in the following and for which the fluctuations at the inflow are continuously increasing. Qualitatively, these results confirm that the present numerical results show a transitional character in the sense that a rather small change in parameters has a large impact on the macroscopic flow behavior such as the jet breakdown location. In order to draw precise conclusions which of the two aspects is more relevant for the observed differences between the experimental results and the present numerical results using the precursor simulation approach, we conclude that more detailed experimental studies, e.g., with a precise characterization of the fluctuation level or with the fluctuations at the inflow reduced to a minimum, would be necessary. Since the uncertainty in the mass flow rate or the Reynolds number has been estimated to~$10\%$ in the experimental studies, it appears to be plausible that less imperfections and disturbances are present in the numerical simulations using the precursor simulation approach than in experimental studies causing a delayed breakdown of the jet as compared to experiments. At the same time, it is surprising that previous LES studies reported excellent agreement with experiments in light of the large sensitivity with respect to certain parameters observed in this work.

\subsection{Reynolds number~$\mathrm{Re}_{\mathrm{th}}=3500$}
\begin{figure}[!ht]
 \centering
 \subfigure[Profile of mean streamwise velocity along centerline.]{
	\includegraphics[width=0.8\textwidth]{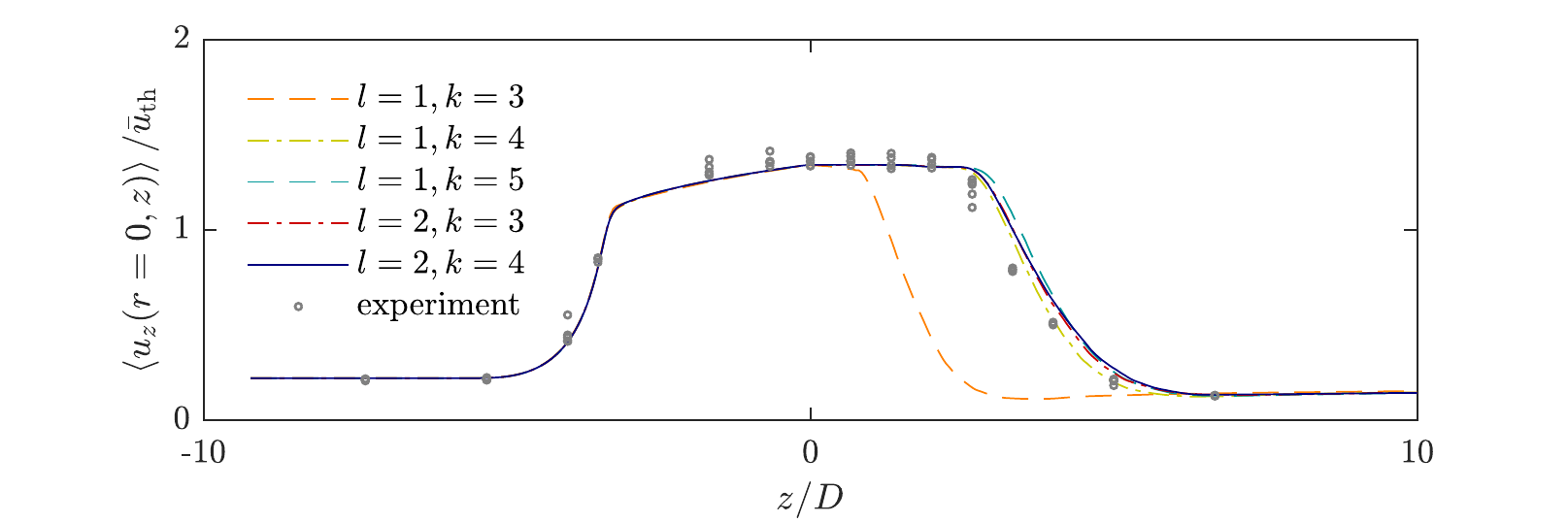}}
 \subfigure[Radial profiles of mean streamwise velocity at various locations~$z_i$,~$i=1,...,12$.]{
	\includegraphics[width=0.8\textwidth]{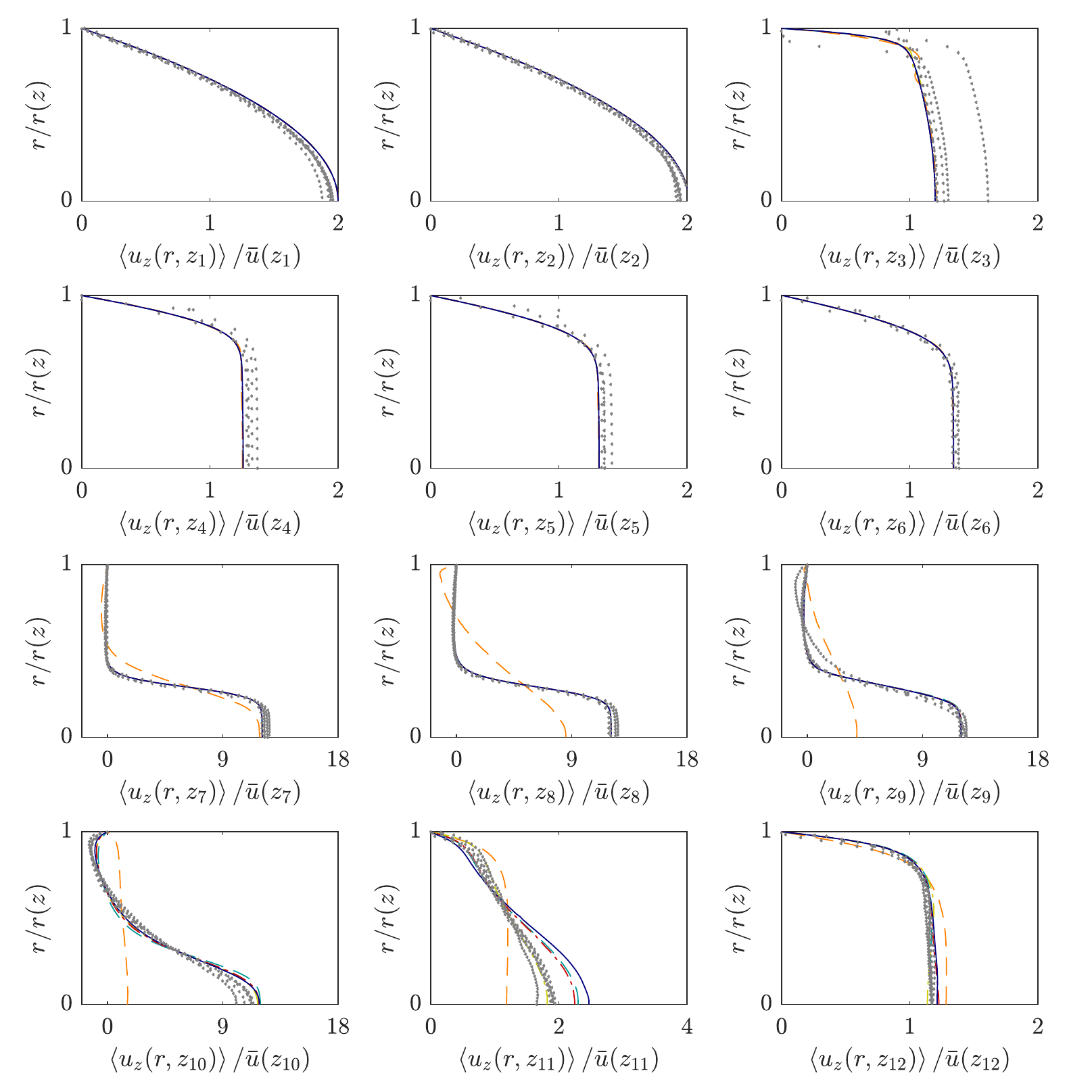}}
\caption{Numerical results for FDA benchmark nozzle model (sudden expansion) at~$\mathrm{Re}_{\mathrm{th}}=3500$.}
\label{fig:results_Re3500}
\end{figure}

In contrast to the previous Reynolds number, the~$\mathrm{Re}_{\mathrm{th}}=3500$ test case is characterized by a turbulent flow at the outflow. This test case has been subject of several LES studies. In~\cite{Passerini2013,Chabannes2017}, good agreement with experimental results has been obtained. In these works, a laminar profile is prescribed at the inflow boundary and only a single spatial resolution is considered without showing mesh convergence results. In~\cite{Zmijanovic2017}, the prediction of the jet breakdown location has been observed to be very sensitive to a change in the parameters of the numerical discretization scheme. As a remedy, the authors of~\cite{Zmijanovic2017} suggest to add turbulent fluctuations to the laminar inflow boundary condition in order to obtain a robust prediction of the jet breakdown location and it has been reported that the amplitude of these perturbations has only a minor influence on the jet breakdown location. Finally, this test case has also been analyzed in~\cite{Delorme2013} but inaccuracies are observed that might be related to an insufficient spatial resolution.

We present mesh convergence results as well as a comparison to experimental results in Figure~\ref{fig:results_Re3500}. Apart from the coarsest mesh, only small differences are observed between the different spatial resolutions and the prediction of the jet breakdown location is in good agreement with experimental studies. For the two finest spatial resolutions, the profiles for the mean streamwise velocity along the centerline almost coincide and the results can be considered as grid-converged. A comparison of radial velocity profiles shows an overall very good agreement with the experiments and small deviations can only be observed at~$z_{10}$ and~$z_{11}$. In the experiments, the jet breaks down slightly earlier as compared to the numerical results. A possible explanation might again be a larger fluctuation level at the inflow boundary. Interestingly, the relatively coarse mesh with~$l=1$ and~$k=4$ exhibits the best agreement with the experimental results, which can be seen from the axial velocity profile as well as the radial velocity profile at~$z_{11}$. This observation highlights the importance to always perform a mesh refinement study. Obviously, it is not guaranteed that the results converge uniformly to the reference solution under mesh refinement. A large sensitivity of the jet breakdown location as reported in~\cite{Zmijanovic2017} where a laminar inflow profile has been used can not be observed. This might be due to the precursor simulation approach suggested in the present work. A more detailed discussion of this aspect can be found in Section~\ref{LaminarInflowVsPrecursor} where results are shown using a laminar velocity profile at the inflow boundary instead of the precursor simulation approach.

\subsection{Reynolds number~$\mathrm{Re}_{\mathrm{th}}=5000$}
\begin{figure}[!ht]
 \centering
 \subfigure[Profile of mean streamwise velocity along centerline.]{
	\includegraphics[width=0.8\textwidth]{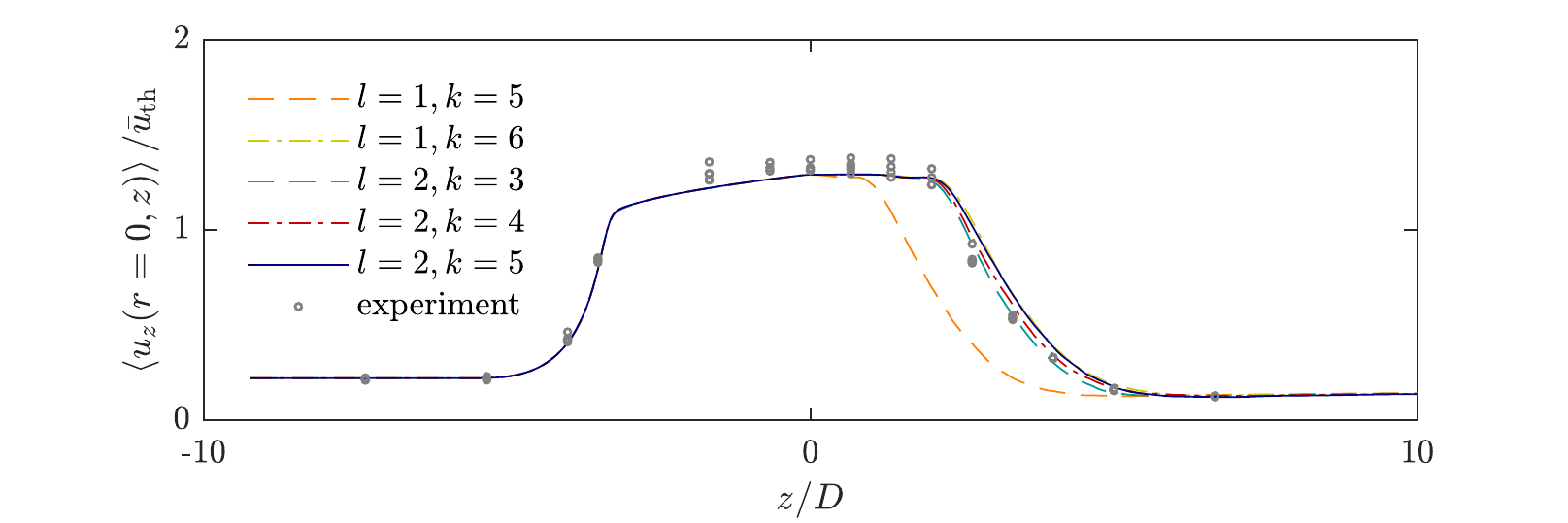}}
 \subfigure[Radial profiles of mean streamwise velocity at various locations~$z_i$,~$i=1,...,12$.]{
	\includegraphics[width=0.8\textwidth]{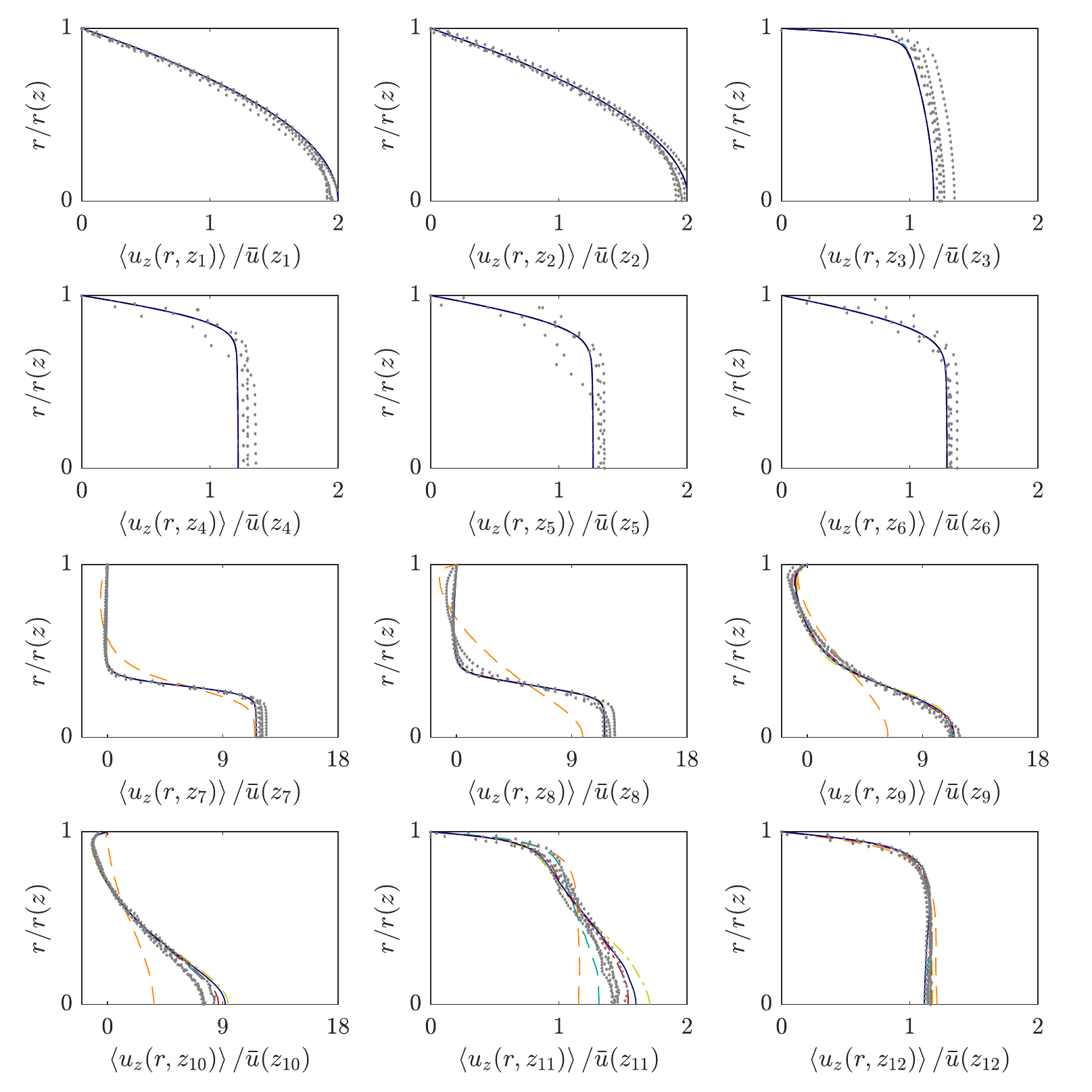}}
\caption{Numerical results for FDA benchmark nozzle model (sudden expansion) at~$\mathrm{Re}_{\mathrm{th}}=5000$.}
\label{fig:results_Re5000}
\end{figure}

In terms of the flow regimes in the inflow and outflow sections, the~$\mathrm{Re}_{\mathrm{th}}=5000$ test case is similar to the previous one at~$\mathrm{Re}_{\mathrm{th}}=3500$. In previous studies using the large-eddy simulation approach, a Reynolds number of~$\mathrm{Re}_{\mathrm{th}}=5000$ has been simulated in~\cite{Delorme2013} where comparably large inaccuracies have been observed as compared to experimental results which are ascribed to an insufficient mesh resolution. Good agreement with experimental results is achieved in~\cite{Nicoud2018}, but results are only shown for a single spatial resolution. Numerical results obtained for several spatial resolutions are shown in Figure~\ref{fig:results_Re5000} for the present discretization approach. An excellent agreement with experimental results is obtained with an accurate prediction of the jet breakdown location. Larger deviations from the experiment are only observed for the coarsest spatial resolution with~$l=1$ and~$k=5$ analyzed here. In the throat section as well as the outflow section close to the sudden expansion the mean streamwise velocity on the centerline is slightly lower than in the experiments. A similar trend has already been observed for~$\mathrm{Re}_{\mathrm{th}}=2000$ and~$\mathrm{Re}_{\mathrm{th}}=3500$.

\subsection{Reynolds number~$\mathrm{Re}_{\mathrm{th}}=6500$}
\begin{figure}[!ht]
 \centering
 \subfigure[Profile of mean streamwise velocity along centerline.]{
	\includegraphics[width=0.8\textwidth]{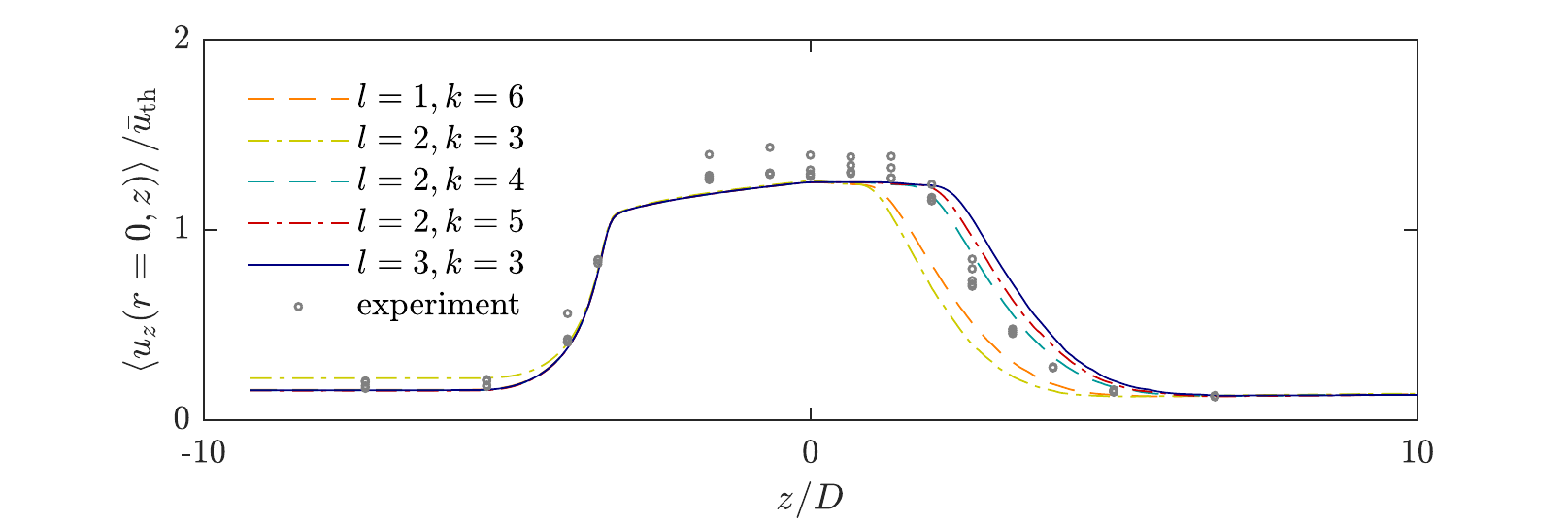}}
 \subfigure[Radial profiles of mean streamwise velocity at various locations~$z_i$,~$i=1,...,12$.]{
	\includegraphics[width=0.8\textwidth]{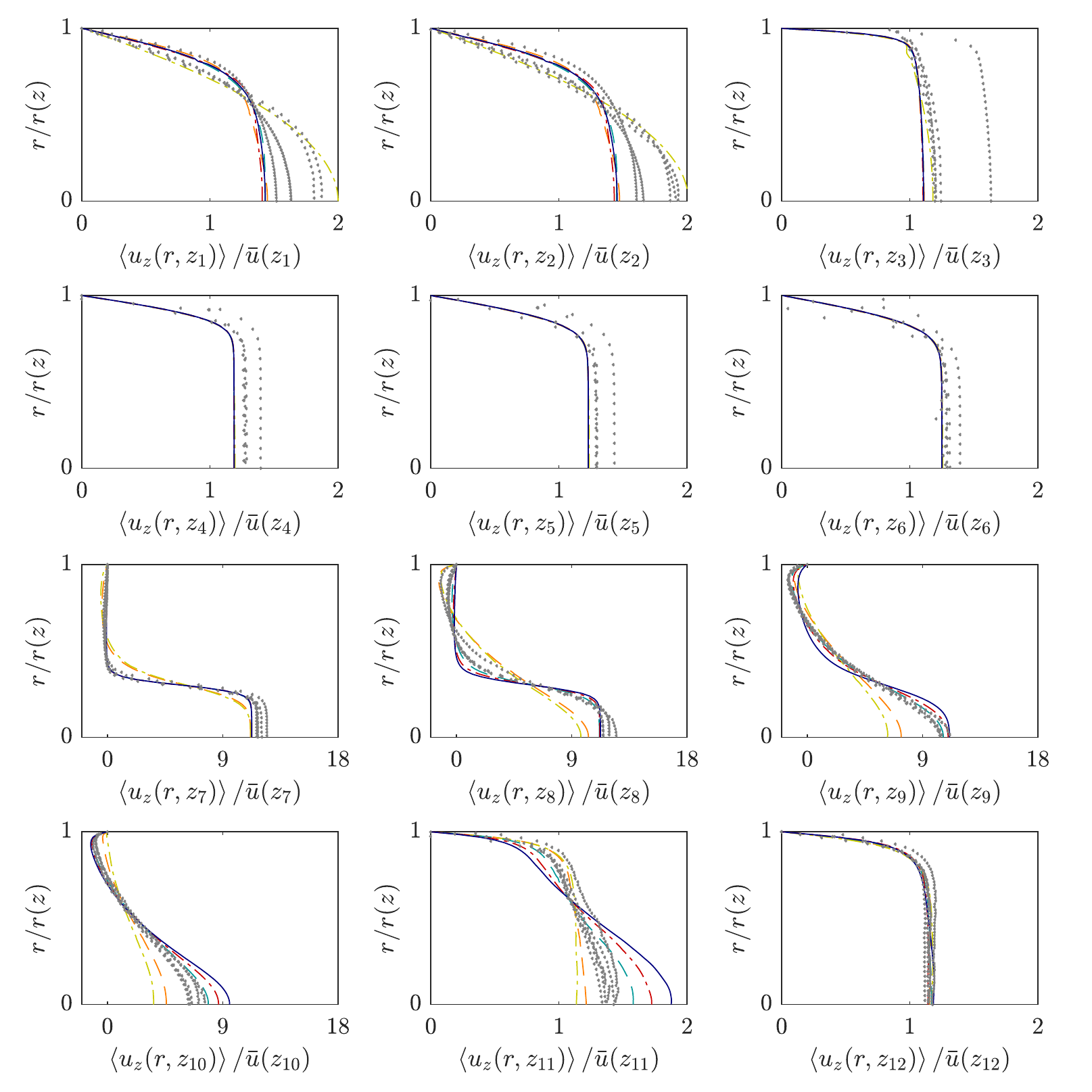}}
\caption{Numerical results for FDA benchmark nozzle model (sudden expansion) at~$\mathrm{Re}_{\mathrm{th}}=6500$.}
\label{fig:results_Re6500}
\end{figure}
For~$\mathrm{Re}_{\mathrm{th}}=6500$, the flow can be expected to be turbulent at the outflow and in the transitional regime at the inflow, see Table~\ref{tab:ReynolsNumbersAndFlowRates}. LES results for the~$\mathrm{Re}_{\mathrm{th}}=6500$ test case have been shown in~\cite{Janiga2014} for one mesh resolution and a mesh convergence study could not be performed due to the large amount of required computational costs. Radial profiles are compared to the experimental results in the outflow section behind the sudden expansion and excellent agreement has been obtained without the need to adapt parameters of the turbulence model. Despite the uncertainty regarding the flow regime at the inflow, which is also confirmed by experimental results~\cite{Hariharan2011} with some experiments showing a laminar profile and others a turbulent one, a laminar inflow profile without turbulent fluctuations has been prescribed. In the present work, the precursor simulation approach is used for all results shown in this section. Numerical results obtained for different spatial resolutions are presented in Figure~\ref{fig:results_Re6500}. By considering the spatial resolutions and the convergence behavior of the results one can observe that the~$\mathrm{Re}_{\mathrm{th}}=6500$ test case is the most challenging one in terms of the required spatial resolution. Even for the finest spatial resolutions investigated in this study, the results for the mean streamwise velocity are not fully grid-converged. On the finest meshes, the jet breakdown location occurs further downstream as compared to the experimental results. While the best agreement with experimental results can be observed for an intermediate resolution with parameters~$l=2$ and~$k=4$, the jet breaks down at larger~$z$-values on the two finest meshes. A qualitatively similar behavior has already been observed for the~$\mathrm{Re}_{\mathrm{th}}=3500$ test case. As a result, demonstrating excellent agreement with experimental results for one specific spatial resolution does not imply that the numerical results are already converged for this specific resolution. Similar to the lower Reynolds number test cases, the mean axial velocity on the centerline is under-predicted in all simulations as compared to the experimental results. However, the results presented here are in agreement with the LES results shown in~\cite{Janiga2014} with respect to this aspect. Regarding the radial velocity profiles, a good agreement with experimental results is achieved. In the inflow section at locations~$z_1$ and~$z_2$, radial profiles show a large variation within the different experiments. In some cases, the inflow profile is laminar, in other cases it is more turbulent which can be explained by the transitional character of the flow at the inlet. This behavior is also reflected in the numerical results. On some meshes, we observed a turbulent flow at the inflow, while the flow becomes laminar for other spatial resolution parameters. This high sensitivity of the inflow profile with respect to the spatial resolution observed for this test case as compared to the lower Reynolds number test cases clearly confirms the transitional character of the flow in the inflow section. We conclude that the present high-order DG approach in combination with the precursor simulation is able to correctly predict the flow behavior in the inflow section. Since the jet breaks down slightly later as compared to the experimental results, the radial velocity profiles at locations~$z_9$,~$z_{10}$, and~$z_{11}$ also deviate from the experimental results for the two finest spatial resolutions. It should be noted, however, that a minor change in the jet breakdown location causes comparably large differences in the radial velocity profiles at these~$z$-locations.

\subsection{Precursor simulation approach versus laminar inflow profile}\label{LaminarInflowVsPrecursor}
To demonstrate the impact of the precursor simulation approach on the numerical results and the prediction of the jet breakdown location, we compare the results to an alternative approach where a parabolic velocity profile is prescribed at the inflow boundary that has been exclusively used in previous LES studies~\cite{Delorme2013,Passerini2013,Janiga2014,Chabannes2017,Zmijanovic2017,Nicoud2018}. In~\cite{Passerini2013}, a large sensitivity of the jet breakdown location has been reported for~$\mathrm{Re}_{\mathrm{th}}=3500$ when using a parabolic inflow profile without additional perturbations and more robust results could be obtained when adding small perturbations to the velocity inflow profile.
\begin{figure}[!ht]
 \centering
\subfigure[Results for~$\mathrm{Re}_{\mathrm{th}}=3500$.]{
 \includegraphics[width=0.8\textwidth]{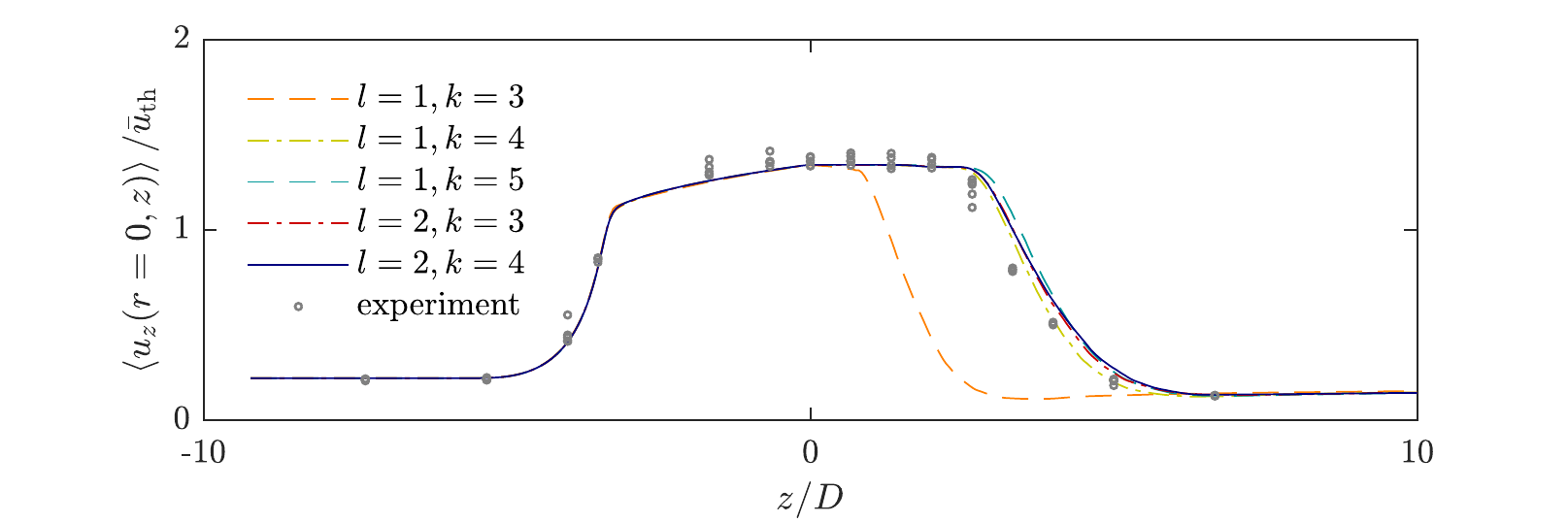}}
\subfigure[Results for~$\mathrm{Re}_{\mathrm{th}}=5000$.]{
 \includegraphics[width=0.8\textwidth]{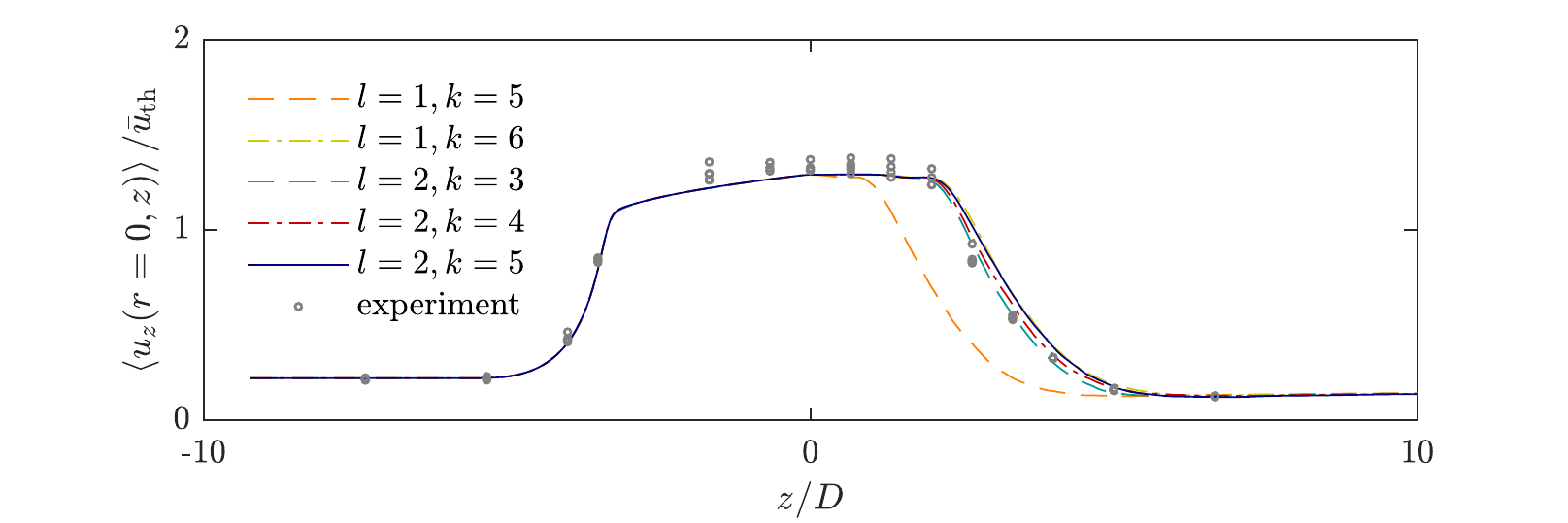}}
\subfigure[Results for~$\mathrm{Re}_{\mathrm{th}}=6500$.]{
 \includegraphics[width=0.8\textwidth]{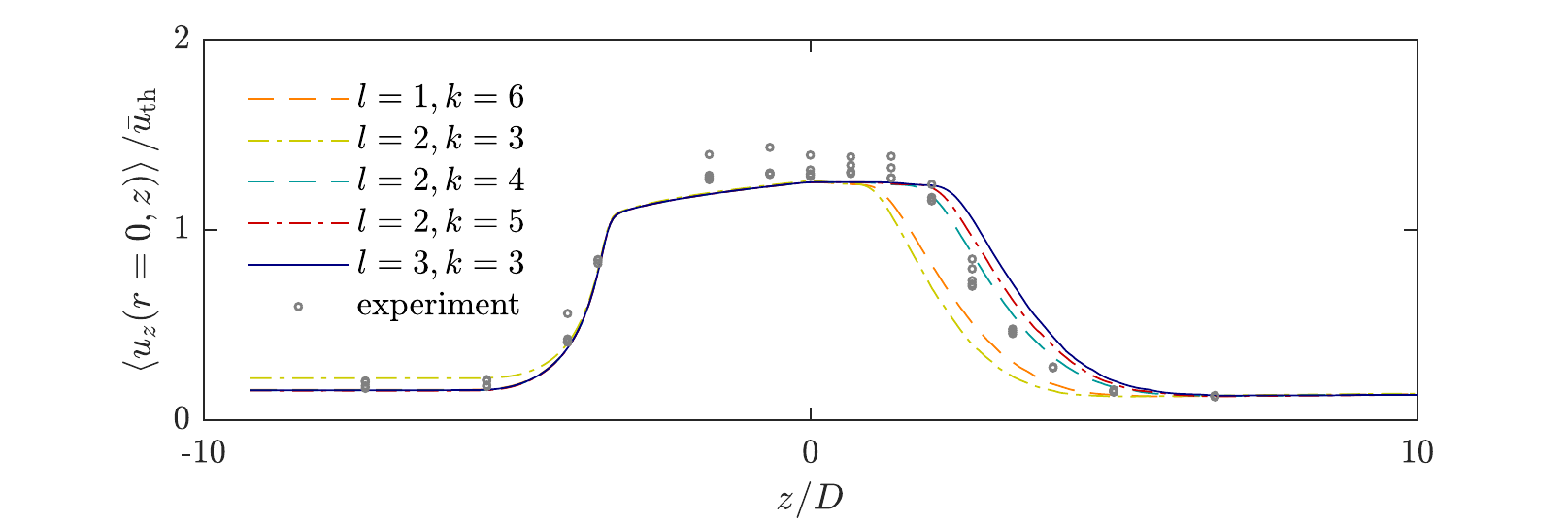}}
\caption{Summary of numerical results for FDA benchmark nozzle model (sudden expansion) for different Reynolds numbers using the precursor simulation approach: Profiles of mean streamwise velocity along centerline.}
\label{fig:summary_results_precursor}
\end{figure}

\begin{figure}[!ht]
 \centering
\subfigure[Results for~$\mathrm{Re}_{\mathrm{th}}=3500$.]{
 \includegraphics[width=0.8\textwidth]{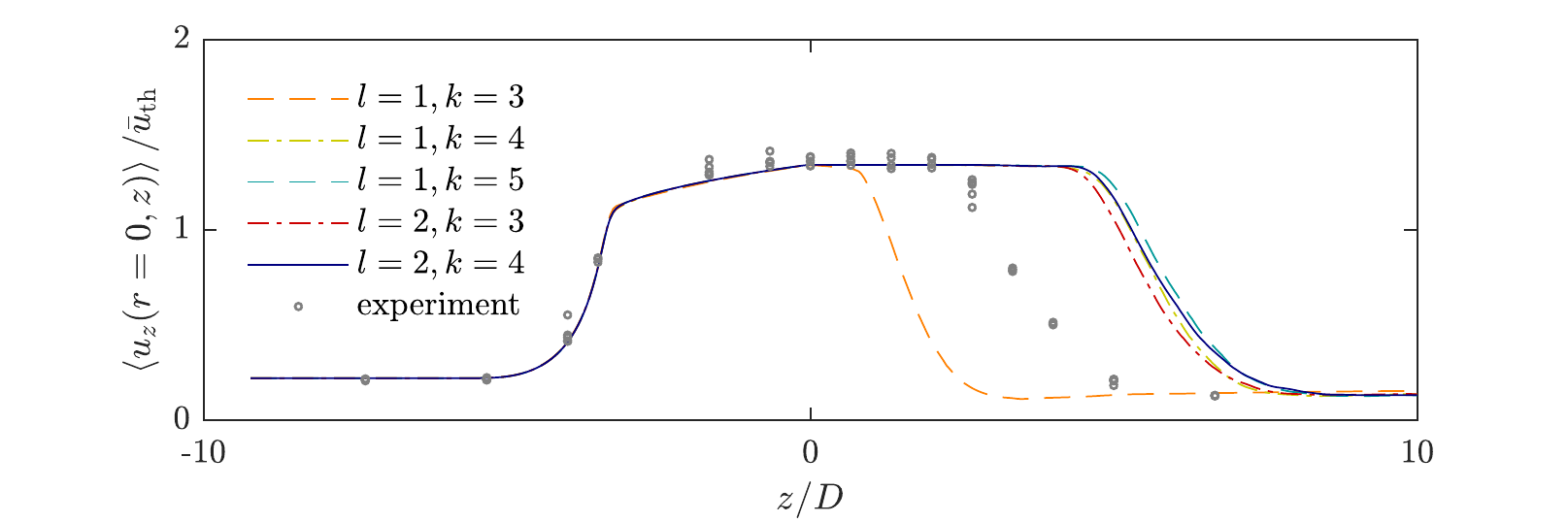}}
\subfigure[Results for~$\mathrm{Re}_{\mathrm{th}}=5000$.]{
 \includegraphics[width=0.8\textwidth]{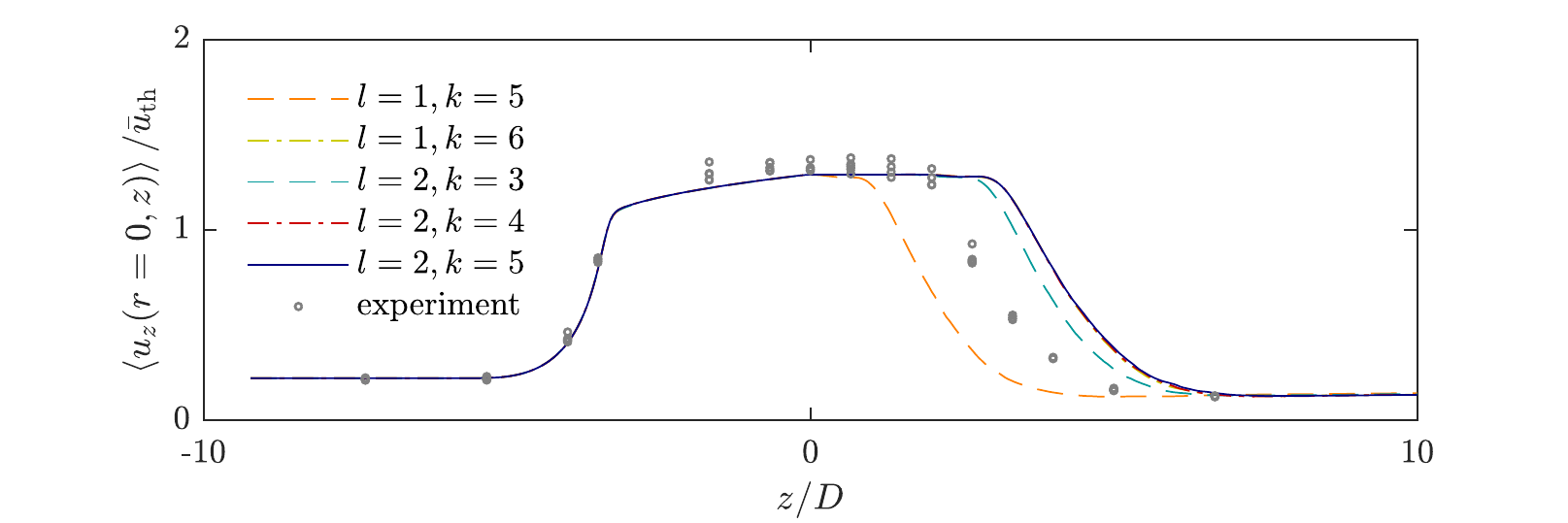}}
\subfigure[Results for~$\mathrm{Re}_{\mathrm{th}}=6500$.]{
 \includegraphics[width=0.8\textwidth]{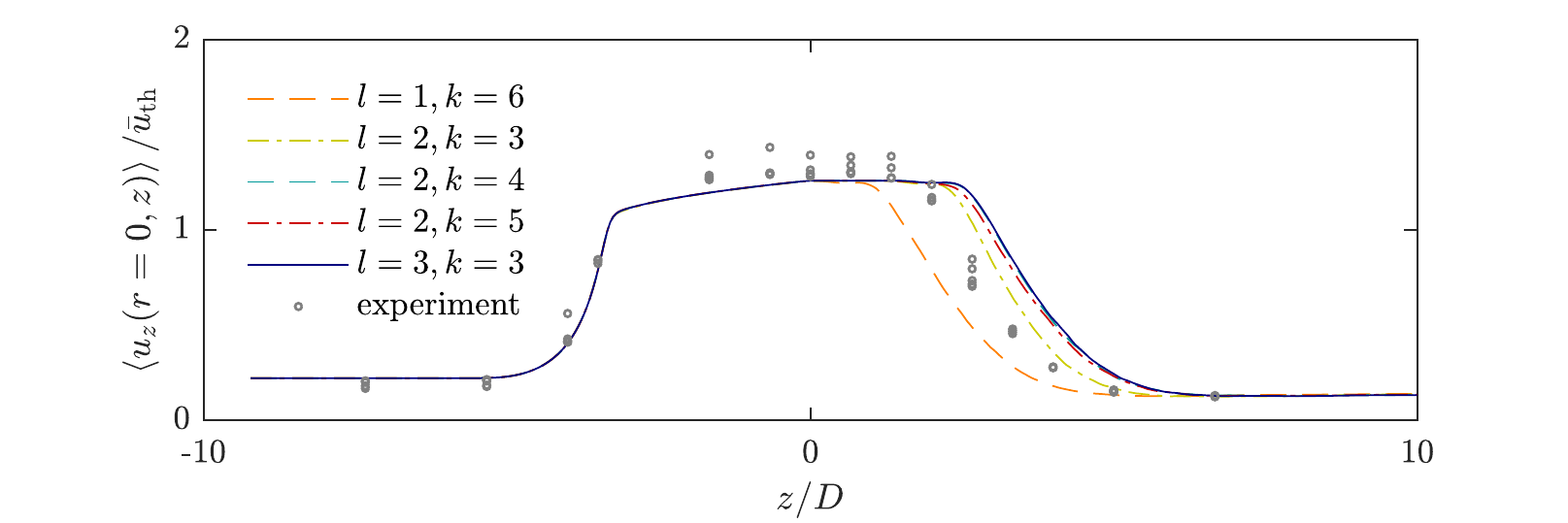}}
\caption{Numerical results for FDA benchmark nozzle model (sudden expansion) for different Reynolds numbers using a parabolic inflow profile instead of the precursor simulation approach: Profiles of mean streamwise velocity along centerline.}
\label{fig:results_laminar_inflow}
\end{figure}

Numerical results shown in previous sections for the precursor approach are summarized in Figure~\ref{fig:summary_results_precursor} for the three turbulent Reynolds numbers~$\mathrm{Re}_{\mathrm{th}}=3500,5000$ and~$6500$ (using the respective data of Figures~\ref{fig:results_Re3500},~\ref{fig:results_Re5000},~\ref{fig:results_Re6500}), and numerical results obtained for a laminar inflow profile are shown in Figure~\ref{fig:results_laminar_inflow} using the same spatial resolutions investigated in the previous sections to allow a direct comparison to the precursor strategy. Since the main aspect of this investigation is the prediction of the jet breakdown location, we only show results for the mean axial velocity profile along the centerline. Similar to the precursor simulation approach, the jet breaks down too early in case of very coarse spatial resolutions. On finer meshes, however, the jet breaks down significantly later as compared to the experimental results. Moreover, the results differ significantly from those obtained with the precursor approach. Especially for~$\mathrm{Re}_{\mathrm{th}}=3500$ and~$5000$,  significant discrepancies can be observed between the precursor approach and the laminar inflow profile with the precursor approach yielding significantly more accurate results with respect to experimental results. The differences are smaller for the highest throat Reynolds number of~$\mathrm{Re}_{\mathrm{th}}=6500$. For both the precursor approach and the laminar inflow profile, the jet breakdown location is slightly larger than in the experiments, but again the precursor approach appears to be more accurate. These results are very interesting and are somehow counter-intuitive. Intuitively, one might argue that the precursor strategy will be more and more important for increasing Reynolds numbers with the flow becoming turbulent at the inflow and that the parabolic inflow profile can be expected to produce accurate results as long as the flow is laminar in the inflow section. Given the fact that the precursor approach predicts a laminar velocity profile at the inflow boundary in agreement with experimental results for~$\mathrm{Re}_{\mathrm{th}}=3500$ and~$5000$ for all spatial resolutions as shown in Figures~\ref{fig:results_Re3500} and~\ref{fig:results_Re5000}, respectively, it might be unexpected that the precursor approach improves the accuracy of the results significantly for these Reynolds numbers. According to our results, the precursor approach seems to introduce fluctuations that allow a correct prediction of the jet breakdown location. This can be seen as a great strength of the percursor approach and an important ingredient towards generic and parameter-free turbulent flow solvers. While the velocity inflow profiles differ considerably for the precursor approach and the parabolic inflow profile at~$\mathrm{Re}_{\mathrm{th}}=6500$, differences in the numerical results are comparably small. Obviously, the jet breakdown location becomes less sensitive to the flow regime at the inflow for this large Reynolds number. The results in Figure~\ref{fig:results_laminar_inflow} show that the numerical results do not converge towards the experimental results under mesh refinement. While it might be possible to find specific spatial resolutions reproducing the jet breakdown location measured in experimental studies, the results should be scrutinized in terms of their reliability. We conclude that an LES study based on a single spatial resolution only can not be considered as reliable given the results shown in this work.

\begin{remark}
The relative computational costs for the precursor simulation do not account for more than approximately~$(40-50)\%$ of the overall costs. At first sight, this is a significant share of the overall computational costs. However, note that increasing the refinement level by~$1$ causes an increase in computational costs by a factor of~$16$ under idealized assumptions (e.g., optimal parallel scalability, mesh-independent iteration counts for linear solvers). Basically, the aim is to achieve efficient numerical methods, but the computational costs alone are not the relevant metric for the efficiency of a numerical method. Instead, efficiency should be defined as the ratio of accuracy and computational costs~\cite{Fehn2018b}. The numerical results shown above with precursor simulation on the one hand and a prescribed laminar inflow profile on the other hand clearly demonstrate that the precursor approach improves the overall efficiency of the method, i.e., the results with precursor simulation for spatial resolution~$l=1$ and~$k=4$ at~$\mathrm{Re}_{\mathrm{th}}=3500$ are computationally cheaper and at the same time more accurate than the computations on the three finest meshes with laminar inflow profile (without precursor domain).
\end{remark}

\section{Conclusion}\label{Conclusion}
In this work, we investigated novel high-order discontinuous Galerkin discretization techniques as an accurate and generic flow solver for the simulation of transitional and turbulent flow problems typical of biomedical applications. As a numerical test case for the validation of the methods, the FDA benchmark nozzle model in sudden expansion configuration has been considered over a wide range of Reynolds numbers involving laminar, transitional, and turbulent flow regimes. By means of a comprehensive numerical investigation including mesh convergence studies for all Reynolds numbers the generality and versatility of the numerical flow solver has been critically assessed. Our main conclusions are the following: The methodology proposed in this work is capable of correctly predicting the flow behavior for all Reynolds numbers. The use of a precursor simulation strategy is a key ingredient towards a generic and parameter-free flow solver. Good or excellent agreement with experimental results has been obtained for~$\mathrm{Re}_{\mathrm{th}}=500,3500,5000,6500$. For the three largest Reynolds numbers, the~$z$-location of the jet breakdown is slightly larger than in the experimental studies. This might be explained by the fact that fluctuations and flow perturbations are larger in the experimental setup. For the transitional test case at~$\mathrm{Re}_{\mathrm{th}}=2000$, our numerical results did not confirm the location of the jet breakdown observed in experimental studies. Possible explanations could be uncertainties in the Reynolds number or disturbances in the inflow section present in the experimental setup. The influence of both parameters has been investigated numerically, showing a large sensitivity of the jet breakdown location for this transitional test case. By the example of the turbulent test cases with throat Reynolds numbers of~$\mathrm{Re}_{\mathrm{th}}=3500,5000,6500$, it has been demonstrated that the use of a parabolic inflow profile as used in previous LES studies leads to significantly less accurate results and that the jet breakdown location does not converge towards the experimental results under mesh refinement. While this problem has been tackled by superimposing turbulent fluctuations at the inflow boundary in previous numerical studies, the present work solves this issue in a generic way by using a precursor simulation strategy. Under these circumstances, investing additional computational effort into a precursor simulation is clearly justified and reasonable. In terms of large-eddy turbulence modeling, an implicit approach without explicit subgrid-scale model has been used. A high-order discontinuous Galerkin discretization approach as used here appears to be highly relevant as a turbulent flow solver in biomedical engineering especially due to the fact that no turbulence model parameters have to be adjusted for this method. Finally, we would like to point to the importance of performing mesh refinement studies no matter how reliable a numerical solution method is.

\appendix

\section*{Acknowledgments}
The research presented in this paper was partly funded by the German Research Foundation (DFG) under the project ``High-order discontinuous Galerkin for the EXA-scale'' (ExaDG) within the priority program ``Software for Exascale Computing'' (SPPEXA), grant agreement no. KR4661/2-1 and WA1521/18-1.

\bibliography{paper}

\end{document}